\documentclass[twocolumn,journal]{IEEEtran}
\usepackage[T1]{fontenc}
\usepackage[latin9]{inputenc}
\usepackage{float}
\usepackage{amstext}
\usepackage{amsmath}
\usepackage{flushend}
\usepackage{enumerate}
\usepackage{xcolor}
\usepackage{stfloats}
\usepackage{subfigure}

\usepackage{array}
\usepackage{multirow}

\usepackage{graphicx}

\usepackage{epstopdf}
\usepackage{amssymb}

\usepackage{amsthm,amsmath,amssymb}

\usepackage{mathrsfs}

\newtheorem{remark}{Remark}

\usepackage{subfigure}

\usepackage{graphics}
\usepackage{diagbox}

\usepackage{makecell}

\usepackage{amssymb}

\usepackage{amsmath}
\usepackage{booktabs}
\usepackage{tabularx}
\usepackage{color}

\usepackage{bm}          
\usepackage{algorithm}
\usepackage{algorithmic} 

\usepackage[unicode=true,
 bookmarks=true,bookmarksnumbered=true,bookmarksopen=true,bookmarksopenlevel=1,
 breaklinks=false,pdfborder={0 0 0},pdfborderstyle={},backref=false,colorlinks=false]
 {hyperref}
\hypersetup{pdftitle={Your Title},
 pdfauthor={Your Name},
 pdfpagelayout=OneColumn, pdfnewwindow=true, pdfstartview=XYZ, plainpages=false}
\usepackage{breakurl}

\usepackage{cite}

\makeatletter

\floatstyle{ruled}
\newfloat{algorithm}{tbp}{loa}
\providecommand{\algorithmname}{Algorithm}
\floatname{algorithm}{\protect\algorithmname}

\makeatother

\begin{document}
\title{Closed-Loop Integrated Sensing, Communication, and Control for Efficient Drone Flight}
\author{Jingli~Li,~\IEEEmembership{Graduate Student Member,~IEEE,}~Yiyan~Ma,~\IEEEmembership{Member,~IEEE,}~Bo~Ai,~\IEEEmembership{Fellow,~IEEE,}\\
Wei~Chen,~\IEEEmembership{Senior Member,~IEEE,}~Weijie~Yuan,~\IEEEmembership{Senior Member,~IEEE,}~Qingqing~Cheng,~\IEEEmembership{Member,~IEEE,}\\
Tongyang~Xu,~\IEEEmembership{Senior Member,~IEEE,}~Guoyu~Ma,~\IEEEmembership{Member,~IEEE,}~Mi~Yang,~\IEEEmembership{Member,~IEEE,}\\
Yunlong~Lu,~\IEEEmembership{Member,~IEEE,}~Wenwei~Yue,~\IEEEmembership{Member,~IEEE,}~Christos~Masouros,~\IEEEmembership{Fellow,~IEEE,}\\
~and~Zhangdui~Zhong,~\IEEEmembership{Fellow,~IEEE}\\

\thanks{Jingli~Li, Yiyan~Ma, Bo~Ai, Wei~Chen, Guoyu~Ma, Mi~Yang, Yunlong~Lu, and Zhangdui~Zhong are with the School of Electronic and Information Engineering, Beijing Jiaotong University, Beijing 100044, China.
Weijie~Yuan is with the School of System Design and Intelligent Manufacturing, Southern University of Science and Technology, Shenzhen 518055, China. 
Tongyang~Xu and Christos~Masouros are with the Department of Electronic and Electrical Engineering, University College London, WC1E 7JE London, U.K.
Qingqing~Cheng is with the School of Electrical Engineering and Robotics, Queensland University of Technology, Brisbane, QLD 4000, Australia. 
Wenwei~Yue is with the State Key Laboratory of Integrated Services Networks, Xidian University, Xi\textquoteright an, Shaanxi 710071, China.\textit{(Corresponding authors: mayiyan@bjtu.edu.cn; boai@bjtu.edu.cn.)}}
}
\maketitle

\begin{abstract}

Low-altitude wireless networks (LAWN) require drones to follow specific trajectories controlled by ground base stations (GBSs). However, given complex low-altitude channel conditions and limited spectrum and power resources, sensing errors and wireless link unreliability cannot be ignored, leading to trajectory deviations that threaten flight safety. To address this issue, this paper proposes an integrated sensing-communication-control (ISCC) closed-loop trajectory tracking approach, aiming to reveal the coupling mechanisms among communication, sensing, and control during drone flight. In detail, we incorporate sensing errors in trajectory state estimation, packet losses in control command transmission, and finite blocklength (FBL) transmission effects into the closed-loop dynamics. First, through theoretical analysis, we identify the dominant role of the time-frequency resources allocated to control in ensuring system stability and derive a lower bound on the resources required to guarantee stable operation. Second, to minimize tracking error, we formulate a time-frequency resource allocation optimization problem for the sensing, communication, and control components, subject to constraints on communication rate and closed-loop stability. Accordingly, a solution algorithm based on successive convex approximation (SCA) is proposed. Third, simulation results indicate that once stability is ensured, system performance is primarily determined by sensing accuracy, with the trajectory tracking error exhibiting an approximately linear dependence on the position error bound. Finally, it is shown that the proposed ISCC scheme avoids trajectory divergence under FBL transmission compared with ISCC designs ignoring control packet loss, and could achieve decimeter-level average tracking accuracy, reducing the error to only 17.37\% of that observed in the baseline global navigation satellite system (GNSS) scheme.

\end{abstract}


\section{Introduction}

Low-altitude wireless networks (LAWN) are widely regarded as a key enabling component of future 6G systems, providing essential support for drones in application scenarios such as urban logistics, infrastructure inspection, environmental monitoring, and emergency response~\cite{yang2024embodied, wu2025low, ai20256g}. In such highly dynamic scenarios, drones are typically required to strictly follow preplanned flight trajectories, which imposes stringent requirements on trajectory tracking accuracy and system stability. However, constrained by complex low-altitude propagation environments and limited wireless resources, sensing errors and wireless link unreliability are inevitable, and can be progressively amplified through closed-loop system dynamics, thereby leading to the persistent accumulation of trajectory deviations and posing potential threats to flight safety.

To meet the requirements of high-precision trajectory tracking, existing studies have primarily focused on in-depth investigations along individual functional dimensions, such as communication, sensing, and control. 
Specifically, sensing-oriented research aims to improve the instantaneous accuracy of localization or state estimation~\cite{yang2018optimal, wang2024spectrum}, communication studies focus on transmission metrics such as throughput and link reliability~\cite{zeng2017energy, wu2018joint}, while control design emphasizes closed-loop stability or the minimization of control costs, for example, the linear quadratic regulator (LQR) cost~\cite{fang2025sensing, han2023r3c}. These efforts have established solid foundations within their respective domains and provided important support for enhancing drone system performance. However, due to the single-functional design perspective, the above studies have not systematically captured the intrinsic interdependencies among sensing, communication, and control, which restricts the overall improvement of trajectory tracking performance.

To overcome the limitations of single-dimension design, integrated sensing, communication, and control (ISCC) has recently attracted significant attention as an emerging cross-layer design framework. 
From a system architecture and scheduling perspective, Chang \emph{et al.}~\cite{chang2022integrated} investigated joint ISCC scheduling mechanisms for cellular-connected drone networks, where sensing information is exploited to assist beam tracking and motion control. From a control-oriented resource optimization perspective, Lei \emph{et al.}~\cite{lei2023control, lei2024edge} proposed a control-oriented power allocation scheme for satellite-drone networks; Jin \emph{et al.}~\cite{jin2025co, jin2025predictive} studied resource allocation and trajectory optimization for cooperative multi-drone systems under the finite blocklength (FBL) transmission regime and within a model predictive control framework, respectively. From a closed-loop modeling and stability analysis perspective, Meng \emph{et al.}~\cite{meng2023modeling, meng2026communication} and Zhou \emph{et al.}~\cite{zhou2025integrated} analyzed the impact of communication imperfections, such as delay and packet loss, on closed-loop stability in industrial wireless networks. These studies characterize the interplay among sensing, communication, and control in ISCC systems from different perspectives.

Essentially, drone trajectory control is a continuous-time dynamic control system that is highly sensitive to latency and accuracy. Under the ISCC architecture, the system state is acquired through real-time ISAC sensing, which inevitably introduces sensing noise and thus leads to observation uncertainty; meanwhile, control commands are delivered over FBL communication links, where potential transmission errors further induce execution uncertainty. These fluctuations in observation accuracy and failures in command execution are coupled within the feedback loop, jointly shaping the stochastic dynamics of the system state evolution. Although the aforementioned ISCC-related studies provide valuable insights, their application to drone trajectory control still faces two critical modeling bottlenecks. On the one hand, existing studies have yet to reveal the intrinsic coupling between the physical-layer characteristics of active ISAC sensing and the closed-loop dynamics. In ISCC closed-loop modeling for LAWN, sensing uncertainty is commonly simplified as additive observation noise with fixed statistical characteristics~\cite{jin2025co}. However, in active ISAC systems, sensing uncertainty is determined by physical-layer resource allocation through its effect on delay-Doppler resolution, and ignoring this dependency makes it difficult to characterize the impact of physical-layer resource allocation on closed-loop control performance.

On the other hand, existing studies have not adequately considered the impact of wireless link unreliability on control command transmission. Most existing studies focus on imperfections in the uplink from sensors to controllers, while implicitly assuming reliable downlink transmission from controllers to actuators~\cite{chang2022integrated, lei2023control, lei2024edge, jin2025co, jin2025predictive, park2017wireless, baillieul2007control}, thereby neglecting the packet loss risk induced by channel fading and FBL effects. As a result, the control input actually applied to the actuator no longer coincides with the intended command, but instead manifests as a random process constrained by the reliability of the downlink. If such stochastic execution deviations and their accumulation effects in closed-loop evolution are not explicitly modeled, the resulting control policy may exhibit a systematic mismatch with the system's actual operational behavior.

To address the lack of a unified modeling approach in existing ISCC studies for characterizing sensing errors and wireless link unreliability in closed-loop evolution, this paper proposes an ISCC closed-loop system model for drone trajectory tracking, aimed at revealing the propagation, coupling, and accumulation mechanisms of full-chain ``sensing-communication-control'' uncertainties. The main contributions of this paper are summarized as follows:

\begin{enumerate}
    \item We develop an ISCC closed-loop system model for drone trajectory tracking in LAWN, which explicitly incorporates ISAC sensing errors, communication data transmission errors, and random control command losses induced by the FBL mechanism into the system state-space equations. This enables unified modeling of sensing, communication, and control uncertainties alongside the drone's dynamic states (position and velocity), facilitating a systematic analysis of the coupling among sensing, communication, and control.

    \item  Theoretical analysis reveals the roles of sensing and control resource fractions in the ISCC closed-loop system. First, the sensing resources suppress observation noise, determining the accuracy of state estimation, which in turn affects the precision of control inputs. Second, the control resources regulate the reliability of control command transmission and thus dominate the mean-square stability of the closed-loop system, based on which a lower bound on the control resources guaranteeing this stability is derived. This analysis quantifies the impact of sensing and control resources on control accuracy and closed-loop stability, respectively.

    \item We formulate a time-frequency resource allocation optimization problem for sensing, communication, and control to minimize the trajectory tracking error, subject to communication rate requirements and closed-loop stability constraints. By exploiting the threshold structure induced by the stability condition to reformulate the feasible region, we develop an efficient successive convex approximation (SCA)-based algorithm to determine the optimal allocation of communication, sensing, and control resources.

    \item Simulation results reveal the intrinsic coupling among communication, sensing, and control performance. Once closed-loop stability is ensured, the overall system performance becomes primarily limited by sensing accuracy, and the trajectory tracking error exhibits an approximately linear relationship with position error bound (PEB). The proposed ISCC closed-loop scheme, under FBL-induced packet losses, is able to suppress control-failure-induced trajectory deviations compared with the ISCC design that ignores control packet loss, and further reduces the average trajectory tracking error to $17.37\%$ of that achieved by the baseline global navigation satellite system (GNSS)-based closed-loop scheme.

\end{enumerate}

The rest of this paper is structured as follows. Section~\ref{System Model} presents the ISCC system model and derives the Cram\'er-Rao lower bound (CRLB) for state estimation. Section~\ref{Close-loop} formulates the closed-loop state-space model, incorporating both sensing errors and stochastic control packet losses. Section~\ref{Minimization} investigates the joint resource allocation problem to minimize trajectory tracking error, solved via a proposed SCA-based algorithm. Section~\ref{Performance Evaluation} provides simulation results to demonstrate the effectiveness of the proposed scheme, and Section~\ref{Conclusion} concludes the paper.

\textbf{Notation:} 
$\|\cdot\|_2$ denotes the Euclidean norm, and $\mathrm{rect}(\cdot)$ the rectangular window function. 
$\mathrm{Tr}(\cdot)$ and $\det(\cdot)$ denote the trace and determinant, respectively. 
$\otimes$ and $\mathrm{vec}(\cdot)$ denote the Kronecker product and column-wise vectorization, while $\rho(\cdot)$ denotes the spectral radius. 
$\Re\{\cdot\}$ denotes the real part. 
$\mathbb{R}^n$ denotes the $n$-dimensional real space, and $\mathbf{I}_n$ and $\mathbf{0}_n$ the $n \times n$ identity and zero matrices. 
$\mathrm{blkdiag}(\cdot)$ denotes the block-diagonal operator. 
$Q^{-1}(\cdot)$ denotes the inverse Gaussian $Q$-function, $h(\cdot)$ the differential entropy, and $N(\cdot)$ the entropy power. 
Finally, $\boldsymbol{\Sigma}_{\mathbf{w}}$ denotes the covariance matrix of $\mathbf{w}$.

\section{System Model}\label{System Model}

In this section, we establish the system model for ISCC-enabled drone tracking, including the closed-loop architecture, the FBL-based communication signal model, the sensing signal model, and derive the CRLB for state estimation.

\subsection{ISCC Closed-Loop Architecture and ISAC Waveform}

\begin{figure}[t]
    \centering
    \includegraphics[width=2.7in]{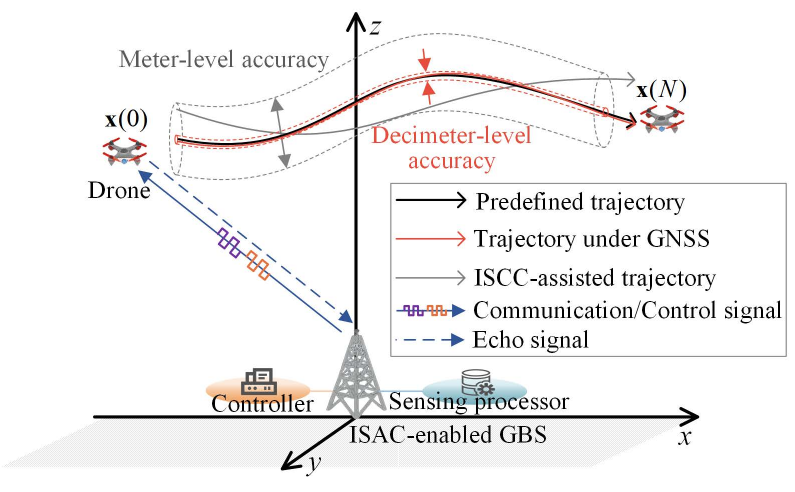}
    \caption{Illustration of ISCC-based drone trajectory tracking.}
    \label{tu1}
\end{figure}

\begin{figure}[t]
    \centering
    \includegraphics[width=2.3in]{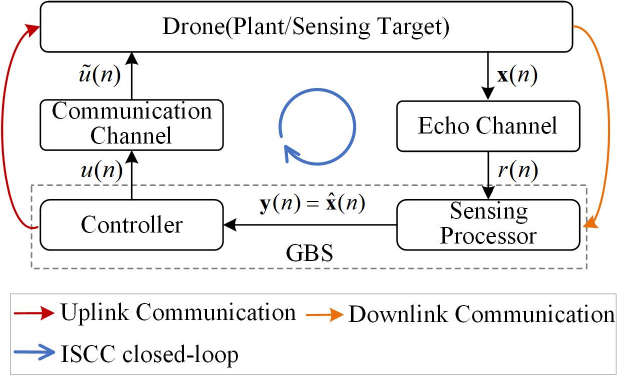}
    \caption{Illustration of the ISAC-enabled ISCC closed-loop operation.}
    \label{tu2}
\end{figure}

We focus on a drone trajectory tracking system that operates under an ISCC mechanism, as shown in Fig.~\ref{tu1}. The system consists of a GBS and a drone\footnote{This paper focuses on the fundamental closed-loop analysis of ISCC in a single-drone setting, with extensions to multi-drone network scenarios left for future work.}, where the GBS integrates an ISAC transceiver, a sensing processor, and a controller to function as a unified ISCC node for drone state estimation and control command generation, while the drone simultaneously serves as a controlled actuator and a sensing target. Within each control cycle, the GBS first senses the drone's state via the ISAC echo, then generates control commands in the controller, which are delivered to the drone through the same dual-function ISAC link. After executing the commands, the drone updates its state, which is subsequently re-sensed by the GBS, forming a continuously operating "sensing-estimation-control-transmission-execution" ISCC closed-loop system, as shown in Fig.~\ref{tu2}. It should be noted that the transmission of control commands constitutes a low-latency, high-reliability communication process, which is distinct from the conventional uplink and downlink data communication links. The system functionality can be categorized into three components: (i) sensing, which performs drone state estimation via the ISAC echo; (ii) data communication, which supports uplink and downlink data exchange between the GBS and the drone; and (iii) control signaling, which enables the GBS to deliver real-time control commands to the drone.

Without loss of generality, a 3D Cartesian coordinate system is established with the GBS located at the origin. The position of the drone at the \(n\)-th time slot is denoted by \(\mathbf{p} = \left[ p_x,\, p_y,\, p_z \right]^{\mathsf T}\), with the corresponding distance, azimuth, and elevation angles relative to the GBS given by \({l} = {\left\| {{\mathbf{p}}} \right\|_2}\), \(\theta = \arctan(p_y/p_x)\), and \(\varphi = \arcsin(p_z/l)\), respectively. We assume that the drone is equipped with a single omnidirectional antenna, while the GBS employs a uniform planar array (UPA) composed of $N_x \times N_y$ antenna elements, where $N_x$ and $N_y$ represent the numbers of elements along the azimuth and elevation directions, respectively. The GBS transmits an orthogonal frequency division multiplexing (OFDM) waveform simultaneously supporting both communication and sensing, with the time-domain signal expressed as~\cite{li2025sensing}
\begin{equation}
\begin{split}
s(t) = & \sum_{m=0}^{{\cal M}-1} \sum_{k=0}^{{\cal K}-1} \sqrt{P_k}\, d_{m,k}\, e^{j2\pi \left(f_c + k\Delta f\right)\left(t - mT_S\right)} \\
& \times \mathrm{rect}\!\left(\frac{t - mT_S}{T}\right),
\end{split}
\end{equation}
where ${\cal M}$ and ${\cal K}$ denote the numbers of OFDM symbols and subcarriers, respectively, and $d_{m,k}$ is a normalized modulation symbol satisfying $\mathbb{E}[|d_{m,k}|^2] = 1$. The total transmit power $P_t$ is uniformly distributed across subcarriers, yielding a per-subcarrier power of $P_k = P_t/{\cal K}$. The carrier frequency and subcarrier spacing are denoted by $f_c$ and $\Delta f$, respectively. The OFDM symbol duration is $T_S = T + T_{\rm CP}$, where $T = 1/\Delta f$ is the effective symbol duration and $T_{\rm CP}$ denotes the cyclic prefix duration. To support ISCC operation, the available time-frequency resources are partitioned among sensing, data transmission, and control command transmission. 
Let $\alpha_{\rm sen}$, $\alpha_{\rm comm}$, and $\alpha_{\rm ctrl}$ denote the corresponding time-frequency resource fractions, satisfying $\sum_{i} \alpha_i = 1$, $i \in \{\mathrm{sen}, \mathrm{comm}, \mathrm{ctrl}\}$. In the OFDM framework, $\alpha_i$ represents the fraction of allocated subcarriers. Accordingly, the effective bandwidth and transmit power assigned to each function are $B_i = \alpha_i B$, $\quad P_i = \alpha_i P_t$, where $B$ denotes the total system bandwidth.

\subsection{Communication Signal Model under FBL}

In the ISCC closed-loop, the communication signals transmitted by the GBS are used to support both communication data transmission and control command delivery. Assuming that the maximum Doppler shift of the communication link satisfies $\nu_C = v_r / \lambda \ll \Delta f$, the frequency-domain response of the communication link at the $m$-th OFDM symbol and the $k$-th subcarrier $\mathbf{H}_{C,m,k} \in \mathbb{C}^{N_x N_y \times 1}$ is defined as~\cite{ma2025delay}
\begin{equation}
\mathbf{H}_{C,m,k} = g_{C,m,k} \sqrt{\beta_C}\, \mathbf{a}(\theta, \varphi)\, e^{-j A_k \tau_C} e^{j B_m \nu_C},
\label{eq:comm_channel}
\end{equation}
where $g_{C,m,k} \sim \mathcal{CN}(\mu_C, \sigma_C^2)$ is the small-scale fading coefficient, with $\mu_C$ and $\sigma_C^2$ denoting the line-of-sight (LoS) and scattering components, respectively. $\tau_C = l/c$ is the communication propagation delay, $l$ is the distance between the drone and the GBS, and $c$ is the speed of light. $A_k \triangleq 2 \pi k \Delta f$ and $B_m \triangleq 2 \pi m T_S$ represent the frequency and time phase terms, respectively. The drone's antenna array response vector is $\mathbf{a}(\theta, \varphi) = \mathbf{a}_x \otimes \mathbf{a}_y$, with the horizontal component $\mathbf{a}_x = [1, e^{j 2\pi \frac{d_\lambda}{\lambda} \sin\theta \cos\varphi}, \ldots, e^{j 2\pi (N_x-1)\frac{d_\lambda}{\lambda} \sin\theta \cos\varphi}]^\mathsf T$ and the vertical component $\mathbf{a}_y = [1, e^{j 2\pi \frac{d_\lambda}{\lambda} \sin\theta \sin\varphi}, \ldots, e^{j 2\pi (N_y-1)\frac{d_\lambda}{\lambda} \sin\theta \sin\varphi}]^\mathsf T$, where $d_\lambda$ denotes the antenna spacing and $\lambda$ is the carrier wavelength. $\nu_C = v_r/\lambda$ is the Doppler shift induced by the relative radial velocity $v_r$ between the drone and the GBS. The large-scale channel gain is modeled as $\sqrt{\beta_C}=10^{-({\cal L}+\varepsilon)/20}$, which accounts for both the path loss ${\cal L}$ and shadow fading $\varepsilon \sim \mathcal{N}(0, \sigma_{\rm SF}^2)$. According to the 3GPP UMa-AV LoS model~\cite{3gpp_tr_36_777}, the path loss (in dB) is given by ${\cal L}(l)=28.0+22\log_{10}(l)+20\log_{10}(f_c)$, where $l$ is measured in meters and the carrier frequency $f_c$ is in GHz.

Given the transmit beamforming vector $\mathbf{w} \in \mathbb{C}^{N_x N_y \times 1}$, the received signal at the drone is expressed as $y_{m,k} = \mathbf{H}_{C,m,k}^{\rm H} \mathbf{w} \sqrt{P_k} d_{m,k} + n_{C,m,k}$, where $n_{C,m,k}\sim\mathcal{CN}(0,N_0B)$ is additive white Gaussian noise (AWGN). The communication beamforming gain is defined as $G(\theta,\varphi)=|\mathbf{a}^{\rm H}(\theta,\varphi)\mathbf{w}|^2$. Defining the communication beamforming gain as $G(\theta,\varphi) = |\mathbf{a}^{\rm H}(\theta,\varphi)\mathbf{w}|^2$, the received signal power is $P_{r} = |g_{C,m,k}|^2 P_k \beta_C G_(\theta,\varphi)$. Accordingly, the received signal-to-noise ratio (SNR) at the drone is given by~\footnote{Under the uniform per-subcarrier power allocation and white-noise assumptions, communication data and control signaling experience identical received SNR, which is independent of the time-frequency resource fraction $\alpha_i$.}
\begin{equation}
\gamma=\frac{|g_{C,m,k}|^2 P_t\beta_C G(\theta,\varphi)}{N_0 B}.
\end{equation}

Considering transmission errors under the FBL regime, the effective transmission rates of communication data and control signaling are given by~\cite{polyanskiy2010channel}
\begin{equation}
\mathcal{R}_i
=
\alpha_i B
\left[
\log_2\!\left(1+\gamma\right)
-
\sqrt{\frac{V}{L_i}}\,
Q^{-1}\!\left(\varepsilon_i\right)
\right], i \in \{\mathrm{comm},\,\mathrm{ctrl}\},
\label{eq:rate_equation}
\end{equation}
where $L_i$ denotes the coded blocklength, with communication data employing long-packet transmission and control signaling adopting short-packet transmission. $\varepsilon_i$ denotes the block error rate (BLER) of the corresponding link, and $V = (\log_2 e)^2 \left[ 1 - \left(1+\gamma\right)^{-2} \right]$ denotes the channel dispersion.

\subsection{Sensing Signal Model and CRLB Derivation}

The sensing signal transmitted by the GBS is reflected by the target drone and received at the GBS for target parameter estimation. 
Following the similar assumption $\nu_S = 2v_r / \lambda \ll \Delta f$ as in~\eqref{eq:comm_channel}, the frequency-domain response 
$\mathbf{H}_{S,m,k} \in \mathbb{C}^{N_x N_y \times N_x N_y}$ of the sensing link at the $m$-th OFDM symbol and the $k$-th subcarrier 
can be expressed as~\cite{ma2025delay, ma2025channel}
\begin{equation}
\mathbf{H}_{S,m,k} = g_{S,m,k} \sqrt{\beta_S} \, \mathbf{a}(\theta, \varphi) \mathbf{a}^{\rm T}(\theta, \varphi) \, e^{-jA_k\tau_S} e^{jB_m \nu_S},
\end{equation}
where $g_{S,m,k} \sim \mathcal{CN}(\mu_S, \sigma_S^2)$ models the Rician small-scale fading of the sensing link, with $\mu_S$ representing the dominant LoS reflection component and $\sigma_S^2$ denoting the scattering-induced fluctuations. $\sqrt{\beta_S} = 10^{-\frac{\mathcal{L}_S + \varepsilon}{20}}$ denotes the sensing path gain, which accounts for both the sensing path loss $\mathcal{L}_S$ and the shadowing component $\varepsilon \sim \mathcal{N}(0, \sigma_{\rm SF}^2)$. $\tau_S = 2l/c$ is the round-trip propagation delay, and $\nu_S = 2v_r/\lambda$ is the round-trip Doppler shift. The sensing path loss $\mathcal{L}_S$ is determined by the round-trip path loss and the radar cross section (RCS) of drone~\cite{zhang2024cluster}, and can be expressed as $\mathcal{L}_S = 2 \mathcal{L} - 10 \log_{10} (\sigma_{\rm RCS}) + 10 \log_{10} \left( \frac{\lambda^2}{4\pi} \right)$, where $\sigma_{\rm RCS}$ is the RCS of the target drone. After standard OFDM demodulation and matched filtering, the echo signal $\mathbf{r}_{S,m,k}\in \mathbb{C}^{N_x N_y \times 1}$ received at the GBS is given by 
\begin{equation}
\begin{split}
\mathbf{r}_{m,k} &= \mathbf{H}_{S,m,k} \mathbf{w} \sqrt{P_k} d_{m,k} + \mathbf{n}_{S,m,k} \\
&= \sqrt{P_k \beta_S} g_{S,m,k} \, \boldsymbol{\eta}(\theta, \varphi) \, d_{m,k} \, e^{-j A_k \tau_S} e^{j B_m \nu_S} + \mathbf{n}_{S,m,k}, 
\end{split}
\label{eq:sensing_vector_model}
\end{equation}
where $\mathbf{n}_{S,m,k} \sim \mathcal{CN}(\mathbf{0}, \sigma^2 \mathbf{I})$ denotes the additive white Gaussian noise vector with variance 
$\sigma^2$ per complex dimension, and $k \in \mathcal{K}_{\rm sen}$ with $\mathcal{K}_{\rm sen}$ representing the set of subcarriers allocated for sensing. The sensing spatial signature vector is defined as $\boldsymbol{\eta}(\theta, \varphi) \triangleq  \mathbf{a}(\theta, \varphi) \left( \mathbf{a}^{\rm T}(\theta, \varphi) \mathbf{w} \right)$, which captures the effective two-way spatial response by accounting for both the complex transmit gain $\mathbf{a}^{\rm T}(\theta, \varphi) \mathbf{w}$ and the receive array manifold $\mathbf{a}(\theta, \varphi)$.

Based on the multipath information contained in the echo signals, the round-trip delay $\hat{\tau}_S$
can be estimated, from which the radial distance is obtained as
$\hat{l}=c\hat{\tau}_S/2$.
Meanwhile, the angle-of-arrival (AoA) estimation yields the azimuth angle
$\hat{\theta}$ and elevation angle $\hat{\varphi}$.
These parameters allow us to construct the three-dimensional (3D) direction
vector $\mathbf{u}(\hat{\theta},\hat{\varphi})
=
[\cos\hat{\varphi}\cos\hat{\theta},\,
\cos\hat{\varphi}\sin\hat{\theta},\,
\sin\hat{\varphi}]^{\mathsf T}$, and subsequently determine the drone position as $\hat{\mathbf{p}}=\hat{l} \cdot \mathbf{u}(\hat{\theta},\hat{\varphi})$. As a result, the drone position is determined by the parameter set
$\{\tau_S,\theta,\varphi\}$, and the position-related parameter vector is
defined as $\boldsymbol{\xi}_\mathbf{p}
= [\tau_S,\;\theta,\;\varphi]^{\mathsf T}$. In addition, the phase evolution of the echo signal along the slow-time dimension contains Doppler information, the Doppler shift $\hat{\nu_S}$ can be estimated, from which the drone's radial velocity is obtained as $\hat{v_r} = \lambda \hat{\nu_S}/2$. To jointly quantify the sensing accuracy of both the drone's position and
velocity, we further define the channel parameter vector as
$\boldsymbol{\xi}
= [\tau_S,\;\theta,\;\varphi,\;\nu_S]^{\mathsf T}$, and the corresponding likelihood function is given by
\begin{equation}
L\big(r \,|\, \boldsymbol{\xi}\big)
=
\prod_{m=0}^{\mathcal{M}-1}
\prod_{k \in \mathcal{K}_{\rm sen}}
\frac{1}{\pi \sigma^2}
\exp\!\left(
-\frac{
\big| r_{m,k} - \mu_{m,k} \big|^2
}{\sigma^2}
\right),
\end{equation}
where $\mu_{m,k}
=
\sqrt{P_k \beta_S} g_{S,m,k} \, \boldsymbol{\eta}(\theta, \varphi) \,
d_{m,k}\,
e^{-j A_k \tau_S}
e^{j B_m \nu_S}$.

By defining the error term as $e_{m,k} = r_{m,k} - \mu_{m,k}$, the corresponding log-likelihood function can be expressed as
\begin{equation}
\ln \Big( L(r \,|\, \boldsymbol{\xi}) \Big) 
= - \mathcal{M} \alpha_{\rm sen}  \mathcal{K} \ln (\pi \sigma^2) 
- \frac{1}{\sigma^2} \sum_{m=0}^{\mathcal{M}-1} 
\sum_{k \in \mathcal{K}_{\rm sen}} | e_{m,k} |^2,
\end{equation}
where $|e_{m,k}|^2 = e_{m,k} \, e_{m,k}^* = \left( {{r_{m,k}} - {\mu _{m,k}}} \right)\left( {r_{m,k}^* - \mu _{m,k}^*} \right)$.

For the complex Gaussian observation model, the elements of the Fisher Information Matrix (FIM) for the parameter vector $\boldsymbol{\xi}$ are given by~\cite{ma2025delay}
\begin{equation}
[\mathbf{J}(\boldsymbol{\xi})]_{i,j}
=
\frac{2}{\sigma^2}
\sum_{m=0}^{\mathcal{M}-1}
\sum_{k \in \mathcal{K}_{\rm sen}}
\Re\!\left[
\frac{\partial \mu_{m,k}}{\partial \xi_i}
\frac{\partial \mu_{m,k}^*}{\partial \xi_j}
\right],
\end{equation}
where $\xi_i,\xi_j \in \{\tau_S, \theta, \varphi, \nu_S\}$. The explicit expressions for each element are given in \eqref{eq:FIM_tau}--\eqref{eq:FIM_zero}.
\begin{subequations}
\begin{align}
&J_{\tau_S \tau_S} = \frac{8 (\pi \Delta f)^2 P_k \beta_S \|\boldsymbol{\eta}(\theta, \varphi)\|^2 \, \mathcal{M} \, (\alpha_{\rm sen} \mathcal{K})^3}{3 \, \sigma^2}, \label{eq:FIM_tau} \\
&J_{\theta \theta} = \frac{P_k \beta_S \|\boldsymbol{\eta}(\theta, \varphi)\|^2 \  \, \mathcal{M} \, \alpha_{\rm sen} \mathcal{K}}{2 \, \sigma^2} 
\left\| \frac{\partial \boldsymbol{\eta}(\theta, \varphi)}{\partial \theta} \right\|^2, \label{eq:FIM_theta} \\
&J_{\varphi \varphi} = \frac{P_k \beta_S  \|\boldsymbol{\eta}(\theta, \varphi)\|^2  \, \mathcal{M} \, \alpha_{\rm sen} \mathcal{K}}{2 \, \sigma^2} 
\left\| \frac{\partial \boldsymbol{\eta}(\theta, \varphi)}{\partial \varphi} \right\|^2, \label{eq:FIM_phi} \\
&J_{\theta \varphi} = J_{\varphi \theta} = \frac{P_k \beta_S \|\boldsymbol{\eta}(\theta, \varphi)\|^2 \ \, \mathcal{M} \, \alpha_{\rm sen} \mathcal{K}}{2 \, \sigma^2} 
\Re\left\{ \frac{\partial \boldsymbol{\eta}^{\rm H}}{\partial \theta} \frac{\partial \boldsymbol{\eta}}{\partial \varphi} \right\}, \label{eq:FIM_theta_phi} \\
&J_{\nu_S\nu_S}
=
\frac{
8 \left( \pi T_S \right)^2
P_k \beta_S \|\boldsymbol{\eta}(\theta, \varphi)\|^2 \ 
\, \alpha_{\rm sen}\, \mathcal{K}\, \mathcal{M}^3
}{ 3 \sigma^2 }, \\
&J_{\tau_S \theta} = J_{\tau_S \varphi} = J_{\theta \tau_S} = J_{\varphi \tau_S} = 0. 
\label{eq:FIM_zero}
\end{align}
\end{subequations}

\begin{figure*}[!t]
\begin{equation}
\mathbf{J}_\mathbf{p} =
\begin{bmatrix}
J_{\tau_S \tau_S} & J_{\tau_S \theta} & J_{\tau_S \varphi} \\
J_{\theta \tau_S} & J_{\theta \theta} & J_{\theta \varphi} \\
J_{\varphi \tau_S} & J_{\varphi \theta} & J_{\varphi \varphi}
\end{bmatrix}
=
\frac{P_k \beta_S \|\boldsymbol{\eta}\|^2 \mathcal{M}\,\alpha_{\rm sen}\,\mathcal{K}}{2\sigma^2}
\;
\begin{bmatrix}
\dfrac{16 (\pi \Delta f)^2 (\alpha_{\rm sen} \mathcal{K})^2}{3} & 0 & 0 \\[4pt]
0 & \left\| \frac{\partial \boldsymbol{\eta}}{\partial \theta} \right\|^2 &
\Re\left\{ \frac{\partial \boldsymbol{\eta}^{\rm H}}{\partial \theta} \frac{\partial \boldsymbol{\eta}}{\partial \varphi} \right\}\\[4pt]
0 & \Re\left\{ \frac{\partial \boldsymbol{\eta}^{\rm H}}{\partial \theta} \frac{\partial \boldsymbol{\eta}}{\partial \varphi} \right\} &
 \left\| \frac{\partial \boldsymbol{\eta}}{\partial \varphi} \right\|^2
\end{bmatrix}.
\label{eq:FIM1}
\end{equation}
\end{figure*}

Since the Doppler parameter $\nu_S$ is decoupled from the position-related parameters $\{\tau_S,\theta,\varphi\}$, the velocity-domain FIM reduces to the scalar $J_{\nu_S\nu_S}$, while the position-domain FIM $\mathbf{J}_{\mathbf{p}} \in \mathbb{R}^{3\times3}$ is given in~\eqref{eq:FIM1}. The CRLB for the drone's radial velocity is given by
\begin{equation}
\mathrm{CRLB}_{v_r}
=
\left( \frac{\lambda}{2} \right)^2
\left( J_{\nu_S\nu_S} \right)^{-1},
\end{equation}
with the corresponding velocity error bound (VEB) defined as $\mathrm{VEB} = \sqrt{\mathrm{CRLB}_{v_r}}$.

\begin{figure*}[!t]
\begin{equation}
\Upsilon_\mathbf{p} = \frac{{\partial \boldsymbol{p} }}{{\partial \boldsymbol{\xi}_\mathbf{p}}} = 
\begin{bmatrix}
\dfrac{\partial p_x}{\partial \tau_S} & \dfrac{\partial p_x}{\partial \theta} & \dfrac{\partial p_x}{\partial \varphi} \\
\dfrac{\partial p_y}{\partial \tau_S} & \dfrac{\partial p_y}{\partial \theta} & \dfrac{\partial p_y}{\partial \varphi} \\
\dfrac{\partial p_z}{\partial \tau_S} & \dfrac{\partial p_z}{\partial \theta} & \dfrac{\partial p_z}{\partial \varphi}
\end{bmatrix}
= \frac{c}{2}
\begin{bmatrix}
\cos\varphi \cos\theta & -\tau_S \cos\varphi \sin\theta & -\tau_S \sin\varphi \cos\theta \\
\cos\varphi \sin\theta & \tau_S \cos\varphi \cos\theta & -\tau_S \sin\varphi \sin\theta \\
\sin\varphi & 0 & \tau_S \cos\varphi
\end{bmatrix}.
\label{eq:Jacobian}
\end{equation}
\hrulefill
\end{figure*} 

After obtaining $\mathbf{J}_\mathbf{p}$, it is mapped to the position domain through the nonlinear transformation with Jacobian $\Upsilon_\mathbf{p}$ shown in~\eqref{eq:Jacobian}. The CRLB for position estimation $\mathrm{CRLB}_\mathbf{p}$ is computed as
\begin{equation}
\mathrm{CRLB}_\mathbf{p}
=
\left(
\boldsymbol{\Upsilon}_\mathbf{p}\,
\mathbf{J}_\mathbf{p}^{-1}\,
\boldsymbol{\Upsilon}_\mathbf{p}^{\mathsf T}
\right)_{3\times 3},
\label{eq:CRLB_pos}
\end{equation}
where $\mathrm{CRLB}_\mathbf{p}$ is symmetric positive definite, and its diagonal entries give the lower bounds on the estimation variances along each position dimension. In the subsequent closed-loop analysis, $\mathrm{CRLB}_\mathbf{p}$ is used together with $\mathrm{CRLB}_{v_r}$ as the observation-noise covariance. To summarize overall localization accuracy, we define the PEB as ${\mathrm{PEB}} = \sqrt{\operatorname{Tr}(\mathrm{CRLB}_\mathbf{p})}$.

\begin{figure}[t]
    \centering
    \subfigure[]{
        \includegraphics[width=0.4\textwidth]{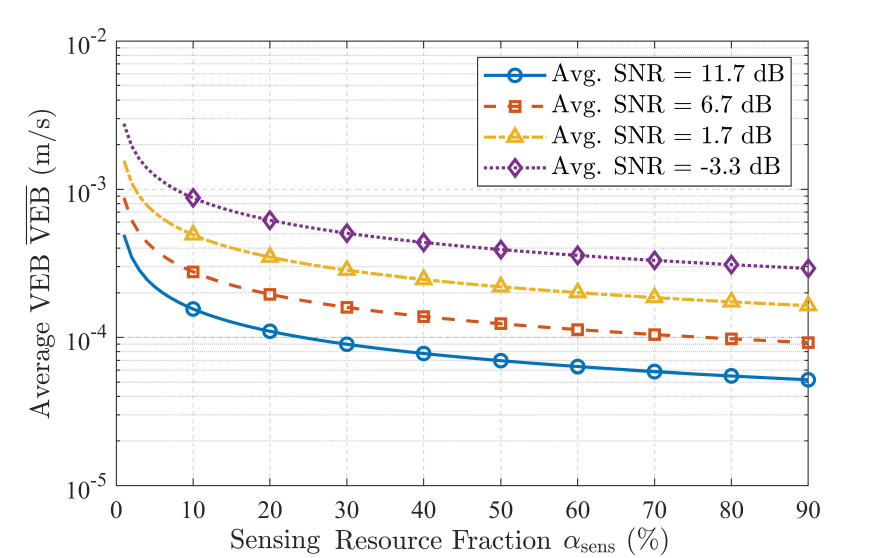}
        \label{fig:subfig4_(a)}
    }
    \hspace{0.5cm}
    \subfigure[]{
        \includegraphics[width=0.4\textwidth]{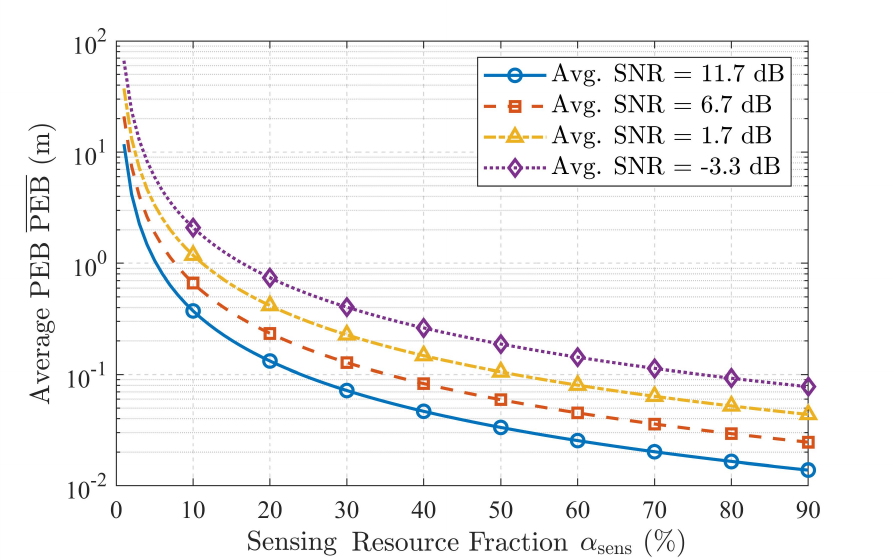}
        \label{fig:subfig4_(b)}
    }
    \caption{(a) Average VEB $\overline{\mathrm{VEB}}$ versus sensing resource fraction $\alpha_{\rm sen}$ for different SNR levels; (b) Average PEB $\overline{\mathrm{PEB}}$ versus sensing resource fraction $\alpha_{\rm sen}$ for different SNR levels.}
    \label{tu4}
\end{figure}

From~\eqref{eq:FIM1}, the sensing resource fraction $\alpha_{\mathrm{sen}}$ determines the number of subcarriers allocated for sensing and thus directly affects the CRLB. Fig.~\ref{tu4} shows the average VEB and PEB as functions of $\alpha_{\mathrm{sen}}$ along the drone trajectory under different SNR conditions. It can be observed that both the VEB and the PEB decrease monotonically with increasing $\alpha_{\mathrm{sen}}$, while the performance gain gradually saturates as $\alpha_{\mathrm{sen}}$ further increases, and this trend remains consistent across different SNR levels. The results indicate that insufficient sensing resource allocation limits the achievable accuracy, whereas excessive allocation yields only marginal performance improvements.

\section{ISCC Close-loop Design}\label{Close-loop}

This section presents the control-oriented design of the ISCC closed-loop system, including a closed-loop model, stability analysis, and a closed-loop performance metric.

\subsection{ISCC Closed-Loop Modeling}

Following the system architecture shown in Fig.~\ref{tu2}, we model the ISCC closed-loop system in a sequential manner, covering the drone dynamics and ISAC-based state observation, the state estimation and control law, and the closed-loop evolution under unreliable control links.

\subsubsection{Drone Dynamics and ISAC-Based Observation}

To characterize the drone flight control dynamics, we adopt a second-order model to describe its 3D motion. The flight duration $T_{\mathrm{m}}$ over the considered trajectory is discretized into $N$ uniform time slots with duration $\Delta t = T_{\mathrm{m}}/N$. With sufficiently small $\Delta t$, the drone dynamics within each slot can be well approximated by a linear model. Moreover, sensing, signal processing, and controller computation delays are neglected, 
as they are typically on the order of microseconds to a few milliseconds and are therefore negligible compared with the second-scale control sampling interval. The system state at time slot $n$ is defined as $\mathbf{x}_n = [\,\mathbf{p}_n^{\mathsf T}\; \mathbf{v}_n^{\mathsf T}\,]^{\mathsf T}
\in \mathbb{R}^6$, with $\mathbf{p}_n \in \mathbb{R}^3$ and $\mathbf{v}_n \in \mathbb{R}^3$ denote the 3D position and 3D velocity of the drone, respectively. The corresponding reference state is given by
$\mathbf{x}_{\mathrm{ref},n}
= [\,\mathbf{p}_{\mathrm{ref},n}^{\mathsf T}\;
\mathbf{v}_{\mathrm{ref},n}^{\mathsf T}\,]^{\mathsf T}$. The discrete-time state evolution of the drone is modeled as~\cite{chang2022integrated}
\begin{equation}
\mathbf{x}_{n+1}
=
\mathbf{A}\mathbf{x}_n
+
\mathbf{B}\tilde{\mathbf{u}}_n
+
\boldsymbol{\omega}_n,
\label{eq:state}
\end{equation}
where $\tilde{\mathbf{u}}_n \in \mathbb{R}^3$ denotes the control input received after FBL transmission, which may differ from the intended control command $\mathbf{u}_n \in \mathbb{R}^3$ due to channel impairments. The system noise is modeled as
$\boldsymbol{\omega}_n \sim \mathcal{N}(\mathbf{0},\boldsymbol{\Sigma}_{\omega})$,
with covariance matrix
$\boldsymbol{\Sigma}_{\omega}
=
\mathrm{blkdiag}(\sigma_\mathbf{p}^{2}\mathbf{I}_3,\;
\sigma_\mathbf{v}^{2}\mathbf{I}_3)$.
The state transition matrix $\mathbf{A}$ and control-input matrix
$\mathbf{B}$ are given by~\cite{aastrom2021feedback}
\begin{equation}
\mathbf{A} =
\begin{bmatrix}
\mathbf{I}_3 & \Delta t\, \mathbf{I}_3 \\
\mathbf{0}_3 & \mathbf{I}_3
\end{bmatrix},
\qquad
\mathbf{B} =
\begin{bmatrix}
\frac{1}{2}\Delta t^{2}\, \mathbf{I}_3 \\
\Delta t\, \mathbf{I}_3
\end{bmatrix}.
\label{eq:AB_matrices}
\end{equation}
where $\mathbf{A}$ and $\mathbf{B}$ are obtained by discretizing the continuous-time second-order kinematic model over the sampling interval $\Delta t$. Under this model, the control input $\tilde{\mathbf{u}}_n$ acts as an equivalent acceleration command applied over one sampling period. For example, given $\mathbf{x}_n = [p_x, p_y, p_z, v, 0, 0]^{\mathsf T}$ and $\tilde{\mathbf{u}}_n = [a, 0, 0]^{\mathsf T}$, the updated state becomes
$\mathbf{x}_{n+1} = [p_x + v\Delta t + \frac{1}{2}a\Delta t^2,\, p_y,\, p_z,\, v + a\Delta t,\, 0,\, 0]^{\mathsf T}$, which follows from Newton's second law.

The GBS acquires noisy drone state observations from ISAC sensing echoes, where the position is observed in three dimensions while the velocity information is limited to the radial component $v_{r,n}$. The observation model is given by
\begin{equation}
\mathbf{y}_n
=
\begin{bmatrix}
\hat{\mathbf{p}}_n^{\mathsf T} &
\hat v_{r,n}
\end{bmatrix}^{\mathsf T}
=
\mathbf{C}_n \mathbf{x}_n
+
\boldsymbol{\upsilon}_n,
\label{eq:observation_model}
\end{equation}
where $\hat{\mathbf{p}}_n$ and $\hat v_{r,n}$ denote the noisy observations of 3D position and radial velocity, respectively. The observation matrix $\mathbf{C}_n$ is given by $\mathbf{C}_n = \mathrm{blkdiag}\!\left(\mathbf{I}_3,\; \mathbf{g}_n^{\mathsf T}\right) \in \mathbb{R}^{4 \times 6}$, where $\mathbf{g}_n$ denotes the unit LoS direction vector used to project the 3D velocity $\mathbf{v}_n$ onto the radial direction, i.e., $v_{r,n} = \mathbf{g}_n^{\mathsf T} \mathbf{v}_n$. The observation noise is modeled as $\boldsymbol{\upsilon}_n \sim \mathcal{CN}(\mathbf{0},\boldsymbol{\Sigma}_{\upsilon,n})$, with $\boldsymbol{\Sigma}_{\upsilon,n}=\mathrm{blkdiag}\!\left(\mathrm{CRLB}_{\mathbf{p},n},\mathrm{CRLB}_{v_r,n}\right)$ determined by the CRLBs of position and radial velocity estimation. 

\subsubsection{State Estimation and Control Law}

Since the velocity information in the observation $\mathbf{y}_n$ is limited to the radial component, a Kalman filter is employed to estimate the full 3D state $\mathbf{x}_n$, with the update rule given by~\cite{haykin2004kalman}
\begin{equation}
\hat{\mathbf{x}}_n = \hat{\mathbf{x}}_{n|n-1} + \mathbf{L}_n (\mathbf{y}_n - \mathbf{C}_n \hat{\mathbf{x}}_{n|n-1}),
\label{eq:KF_update}
\end{equation}
where $\hat{\mathbf{x}}_{n|n-1} = \mathbf{A}\hat{\mathbf{x}}_{n-1} + \mathbf{B}\tilde{\mathbf{u}}_{n-1}$ denotes the predicted state, and $\mathbf{L}_n \in \mathbb{R}^{6 \times 4}$ is the Kalman gain matrix.
By exploiting the system dynamics, the Kalman filter fuses time-series observations with only radial velocity measurements to infer the full state $\hat{\mathbf{x}}_n$, including the 3D velocity $\mathbf{v}_n$.

Based on the estimated state $\hat{\mathbf{x}}_n$, the GBS computes the control command $\mathbf{u}_n$ according to the linear quadratic gaussian (LQG) law~\cite{bertsekas2012dynamic}.
\begin{equation}
\mathbf{u}_n = -\,\mathbf{K}\left( \hat{\mathbf{x}}_n - \mathbf{x}_{\mathrm{ref},n} \right),
\label{eq:LQG_control_input}
\end{equation}
where $\mathbf{K}
= \left( \mathbf{R} + \mathbf{B}^{\mathsf T}\mathbf{P}\mathbf{B} \right)^{-1}
\mathbf{B}^{\mathsf T}\mathbf{P}\mathbf{A} \in \mathbb{R}^{3 \times 6}$ is the feedback gain matrix, which maps the current state deviation to the control input, balancing trajectory tracking accuracy and control effort. Here, $\mathbf{R} \succ 0$ denotes the control weighting matrix, and $\mathbf{P}$ is the optimal cost matrix obtained by solving the discrete algebraic riccati equation: $\mathbf{P} = \mathbf{A}^{\mathsf T}\mathbf{P}\mathbf{A} - \mathbf{A}^{\mathsf T}\mathbf{P}\mathbf{B} (\mathbf{R} + \mathbf{B}^{\mathsf T}\mathbf{P}\mathbf{B})^{-1} \mathbf{B}^{\mathsf T}\mathbf{P}\mathbf{A} + \mathbf{Q}$, where $\mathbf{Q} = \mathrm{blkdiag}\!\left( \mathbf{Q}_\mathbf{p} \mathbf{I}_3,\; \mathbf{Q}_\mathbf{v} \mathbf{I}_3 \right) \succeq 0$ is the state weighting matrix, where $\mathbf{Q}_\mathbf{p}$ and $\mathbf{Q}_\mathbf{v}$ are weighting factors specifically for position and velocity tracking errors, respectively~\cite{bertsekas2012dynamic}.

\subsubsection{Closed-Loop Evolution under Unreliable Control Links}

Due to the stringent latency requirement of control signaling, the control command $\mathbf{u}_n$ must be delivered within a single slot, under which FBL transmission yields a non-zero BLER, leading to random command losses. Consequently, control command transmission is modeled as a Bernoulli packet-drop channel~\cite{meng2026communication}
\begin{equation}
\tilde{\mathbf{u}}_n = \delta_n\, \mathbf{u}_n,
\label{eq:bernoulli_control_channel}
\end{equation}
where $\delta_n$ denotes the packet reception indicator, taking the value $1$ with probability $1-\varepsilon_{{\rm ctrl},n}$ (successful reception) and $0$ with probability $\varepsilon_{{\rm ctrl},n}$ (packet drop). To determine the drop rate $\varepsilon_{{\rm ctrl},n}$, consider a control command of length $S$ bits that must be delivered within the slot duration $\Delta t$, which requires the coding rate to satisfy $\mathcal{R}_{{\rm ctrl}, n} \ge S / \Delta t$. Substituting this rate constraint into the FBL capacity expression~\eqref{eq:rate_equation} yields the packet-drop probability under the latency constraint:
\begin{equation}
\varepsilon_{{\rm ctrl},n}
\le
Q\!\left(
\sqrt{\frac{L_{\rm ctrl}}{V_{\rm ctrl}( \gamma_n )}}
\left[
\log_2(1+\gamma_n)
- \frac{S}{\Delta t\,\alpha_{\rm ctrl} B}
\right]
\right).
\label{eq:ctrl_bler}
\end{equation}

By substituting the observation model~\eqref{eq:observation_model}, the LQG control law~\eqref{eq:LQG_control_input}, and the Bernoulli packet-drop channel~\eqref{eq:bernoulli_control_channel} into state evolution~\eqref{eq:state}, the unified closed-loop state evolution~\eqref{eq:state} is expressed as
\begin{equation}
\mathbf{x}_{n+1}
= \left( \mathbf{A} - \delta_n \mathbf{B}\mathbf{K} \right)\mathbf{x}_n
+ \delta_n \mathbf{B}\mathbf{K}\, \mathbf{x}_{\mathrm{ref},n}
- \delta_n \mathbf{B}\mathbf{K}\, \mathbf{e}_n
+ \boldsymbol{\omega}_n,
\label{eq:closed_loop}
\end{equation}
where $\mathbf{e}_n \triangleq \mathbf{x}_n - \hat{\mathbf{x}}_n$ denotes the state estimation error. For analytical convenience, we define the equivalent process noise as $\tilde{\boldsymbol{\omega}}_n \triangleq \boldsymbol{\omega}_n - \delta_n \mathbf{B}\mathbf{K}\mathbf{e}_n$, whose expected covariance is
\begin{equation}
\Sigma_{\tilde{\omega}}
= \mathbb{E}\!\left[\tilde{\boldsymbol{\omega}}_n 
\tilde{\boldsymbol{\omega}}_n^{\mathsf T}\right]
= \Sigma_{\omega}
+ (1-\varepsilon_{\rm ctrl})\,
\mathbf{B}\mathbf{K}\boldsymbol{\Sigma}_{\mathrm{est}}\mathbf{K}^{\mathsf T}\mathbf{B}^{\mathsf T},
\label{eq:omega_tilde_cov}
\end{equation}
where $\varepsilon_{\mathrm{ctrl}} = \frac{1}{N} \sum_{n=0}^{N-1} \varepsilon_{\mathrm{ctrl},n}$ denotes the average control error. $\boldsymbol{\Sigma}_{\mathrm{est}} \in \mathbb{R}^{6 \times 6}$ is the steady-state estimation error covariance of the Kalman filter, which is determined by the sensing noise covariance $\boldsymbol{\Sigma}_{\upsilon}$.

\begin{remark}
\eqref{eq:closed_loop} explicitly incorporates uncertainties from both the sensing and communication processes into the system state evolution, capturing the parameterized coupling between sensing accuracy, communication reliability, and control dynamics. Specifically, the sensing resource ratio $\alpha_{\rm sen}$ determines the observation noise covariance $\boldsymbol{\Sigma}_v$ via the CRLB, which propagates through the Kalman filter to affect the estimated state $\hat{\mathbf{x}}_n$, and enters the state update ${\mathbf{x}}_{n+1}$ through the feedback gain $K$, thereby directly embedding the impact of sensing accuracy into the closed-loop dynamics. Meanwhile, the control resource ratio $\alpha_{\rm ctrl}$ determines the packet loss probability of control commands under the FBL transmission, which, through the Bernoulli variable $\delta_n$, acts on the closed-loop state transition matrix $(A-\delta_n BK)$, causing communication reliability to directly influence the stochastic evolution of the system state.
\end{remark}

\subsection{Closed-Loop System Stability Analysis}

For the closed-loop system under stochastic control command delivery, we focus on mean-square stability, which provides a stringent guarantee on trajectory tracking performance. Define the state covariance matrix as $\boldsymbol{\Sigma}_n \triangleq \mathbb{E}[\mathbf{x}_n \mathbf{x}_n^{\mathsf T}]$, whose recursion under the closed-loop dynamics~\eqref{eq:closed_loop} is given by
\begin{equation}
\begin{aligned}
\boldsymbol{\Sigma}_{n+1} = &\ (1-\varepsilon_{\rm ctrl})(\mathbf{A}-\mathbf{B}\mathbf{K})\,\boldsymbol{\Sigma}_n\,(\mathbf{A}-\mathbf{B}\mathbf{K})^{\mathsf T} \\
&\ + \varepsilon_{\rm ctrl}\,\mathbf{A}\boldsymbol{\Sigma}_n\mathbf{A}^{\mathsf T} + \boldsymbol{\Sigma}_{\tilde{\omega}}.
\end{aligned}
\label{eq:covariance_recursion}
\end{equation}

To analyze stability, we vectorize the covariance matrix as $\mathbf{s}_n \triangleq \mathrm{vec}(\boldsymbol{\Sigma}_n)$, which leads to
\begin{equation}
\mathbf{s}_{n+1} = \mathbf{M}\,\mathbf{s}_n + \mathrm{vec}(\boldsymbol{\Sigma}_{\tilde{\omega}}),
\label{eq:vec_recursion}
\end{equation}
where the stability matrix is
\begin{equation}
\mathbf{M} = (1-\varepsilon_{\rm ctrl}) \big[(\mathbf{A}-\mathbf{B}\mathbf{K})\otimes(\mathbf{A}-\mathbf{B}\mathbf{K})\big] + \varepsilon_{\rm ctrl}(\mathbf{A}\otimes\mathbf{A}).
\label{eq:M}
\end{equation}

The system is mean-square stable if and only if the spectral radius satisfies $\rho(\mathbf{M}) < 1$. 
Under this condition, the covariance converges to a steady-state value given by $\mathrm{vec}(\boldsymbol{\Sigma}_{\infty}) = \big((\mathbf{I}-\mathbf{M})^{-1}\mathrm{vec}(\boldsymbol{\Sigma}_{\tilde{\omega}})\big)$, ensuring that the drone maintains bounded trajectory deviations in the presence of random packet losses.

\begin{remark}\label{rem:ctrl_pilot_stability}
According to~\eqref{eq:M}, the mean-square stability of the ISCC closed-loop system depends on the system dynamics $\mathbf{A},\mathbf{B}$, the control law $\mathbf{K}$, and the control packet loss rate $\varepsilon_{\rm ctrl}$. Further, from~\eqref{eq:ctrl_bler}, increasing the control resource fraction $\alpha_{\rm ctrl}$ monotonically reduces $\varepsilon_{\rm ctrl}$, which decreases the spectral radius of the state covariance evolution matrix. Therefore, there exists a control resource fraction threshold that guarantees mean-square stability of the ISCC closed loop.
\end{remark}

\subsection{Closed-Loop Performance Metric: LQG Cost}

To evaluate the ISCC closed-loop performance, we adopt the infinite-horizon linear quadratic Gaussian (LQG) cost $b$, which reflects both trajectory tracking errors and control effort while explicitly accounting for ISAC sensing errors $\boldsymbol{\Sigma}_{\mathrm{est}}$ and FBL-induced control packet losses $\varepsilon_{\rm ctrl}$, rather than assuming perfect state information as in the LQR formulation. The cost converges only if the closed-loop system is mean-square stable, i.e., $\rho(\mathbf{M})<1$ in~\eqref{eq:M}, and the LQG cost is defined as~\cite{kostina2019rate}
\begin{equation}
b = \limsup_{N \to \infty} \frac{1}{N}
\sum_{n=1}^{N}
\mathbb{E}\!\left[
\mathbf{x}_n^{\mathsf T} \mathbf{Q} \mathbf{x}_n
+
\tilde{\mathbf{u}}_n^{\mathsf T} \mathbf{R} \tilde{\mathbf{u}}_n
\right].
\label{eq:LQG_cost}
\end{equation}
where $\mathbf{x}_n^{\mathsf T} \mathbf{Q} \mathbf{x}_n$ penalizes the quadratic tracking error from the reference trajectory, while $\tilde{\mathbf{u}}_n^{\mathsf T} \mathbf{R} \tilde{\mathbf{u}}_n$ accounts for the control energy consumption. A smaller value of $b$ indicates better control performance, and for drone flight missions, it means the drone follows the predefined trajectory more closely. Under FBL transmission and random packet drops, achieving a desired performance level $b$ requires the communication link to satisfy the following rate-cost constraint
\begin{equation}
\big(1 - \varepsilon_{\mathrm{ctrl}}\big)\,
\mathcal{R}_{\mathrm{ctrl}}
\ge
L(b),
\label{eq:rate_cost_constraint}
\end{equation}
where ${{\mathcal R}_{\mathrm{ctrl}}} = \frac{1}{N}\sum\nolimits_{n = 0}^{N - 1} {{{\mathcal R}_{\mathrm{ctrl},n}}}$ denotes the average effective transmission rate, and $L(b)$ denotes the rate-cost function that characterizes the minimum average information rate required to maintain the steady-state control cost at $b$, is given by~\cite{jin2025co}
\begin{equation}
L(b)
=
\log_{2}\!\left|\det \mathbf{A}\right|
+
\frac{6}{2}
\log_{2}\!\left(
1+
\frac{
6 \, N(\tilde{\omega}) \,
|\det \mathbf{D}|^{1/6}
}{
b - b_{\min}
}
\right),
\label{eq:Lb}
\end{equation}
where $6$ corresponds to the dimension of the system state vector $\mathbf{x}_n$. $N(\tilde{\omega})
=
\frac{1}{2\pi e}
\exp\!\left(
\frac{2}{6}\,
h(\tilde{\omega})
\right)$ denotes the entropy power of the equivalent noise. $b_{\min}$ denotes the system performance floor, i.e., the minimum achievable control cost under perfect state observation and zero packet loss. \(\mathbf{D}\) captures the effect of state estimation errors and control command losses on the control objective, and is computed together with \(\mathbf{S}\) through the associated Riccati equations~\cite{han2023r3c}
\begin{equation}
\mathbf{D}
=
\big(1 - \varepsilon_{\mathrm{ctrl}}\big)\,
\mathbf{S} \mathbf{B}
\left(\mathbf{R} + \mathbf{B}^{\mathsf T} \mathbf{S} \mathbf{B}\right)^{-1}
\mathbf{B}^{\mathsf T} \mathbf{S}.
\label{eq:Riccati_D}
\end{equation}
\begin{equation}
\mathbf{S}
=
\mathbf{Q}
+
\mathbf{A}^{\mathsf T}
(\mathbf{S} - \mathbf{D})
\mathbf{A}.
\label{eq:Riccati_S}
\end{equation}

The rate-cost constraint in~\eqref{eq:rate_cost_constraint} specifies the information condition required to mitigate uncertainties from the sensing and transmission processes, where the right-hand side $L(b)$ in~\eqref{eq:Lb} gives the minimum information demand, and the left-hand side represents the effective information supply under FBL transmission.

\section{Trajectory Tracking Error Minimization}\label{Minimization}

In this section, we formulate the trajectory tracking error minimization problem, analyze its non-convexity, and develop an SCA-based resource allocation algorithm to efficiently obtain a solution.

\subsection{Problem Formulation}\label{Problem Formulation}

In the ISCC closed-loop system, sensing, communication, and control signaling share a common pool of time-frequency resources. The resource share allocated to sensing determines the accuracy of state estimation, the resource devoted to control affects command reliability and closed-loop stability, while the resource allocated to the communication link must satisfy basic data service requirements. As a result, joint resource allocation fundamentally shapes the closed-loop system dynamics and the achievable trajectory-tracking performance due to the intrinsic coupling between bandwidth and power. However, allocating additional resources to any single function inevitably degrades the others, and insufficient control or sensing resources may even violate closed-loop stability conditions. This motivates a resource allocation design that explicitly accounts for the intrinsic coupling among sensing, communication, and control.

Based on the ISCC closed-loop model established in Section~\ref{Close-loop}, we formulate a joint sensing, communication, and control resource allocation problem aimed at minimizing the time-averaged trajectory tracking error over the mission horizon. Let the reference position at time slot $n$ be
$\mathbf{p}_n^{\mathrm{ref}} = 
\begin{bmatrix}
p_{x,n}^{\mathrm{ref}},\, p_{y,n}^{\mathrm{ref}},\, p_{z,n}^{\mathrm{ref}}
\end{bmatrix}^{\mathsf T}$,
and the actual drone position be $\mathbf{p}_n$. 
The position tracking error is defined as $\mathbf{e}_{\mathbf{p},n} = \mathbf{p}_n - \mathbf{p}_n^{\mathrm{ref}}$. Under the minimum mean-square error (MMSE) criterion, the trajectory error minimization problem is formulated as the joint resource allocation problem, which is given by
\begin{subequations}
\label{prob:resource_allocation}
\begin{align}
\min_{\alpha_{\rm sen},\,\alpha_{\rm ctrl}, \,\alpha_{\rm comm}} \quad
& \mathbb{E}\!\left[
\frac{1}{N}
\sum_{n=0}^{N-1}
\left\lVert \mathbf{e}_{\mathbf{p},n} \right\rVert^2
\right] \label{prob:obj} \\
\text{s.t.}\quad\quad\quad 
& \rho(\mathbf{M}) < 1, \label{cons:stability} \\
& (1 - \varepsilon_{\rm ctrl})\,\mathcal{R}_{\rm ctrl}
\ge L(b), \label{cons:rate_cost} \\
& \mathcal{R}_{\rm comm,n}
\ge \mathcal{R}_{\rm comm}^{\min}, \label{cons:comm_rate} \\
& \alpha_{\rm comm}
+ \alpha_{\rm sen}
+ \alpha_{\rm ctrl}
= 1, \label{cons:resource_sum} \\
& 0 < \alpha_i < 1,\;
i \in \{\rm sen, comm, ctrl\}. \label{cons:alpha_range} 
\end{align}
\end{subequations}
where \eqref{cons:stability} enforces mean-square stability of the closed-loop dynamics under random control packet drops, forming a fundamental feasibility condition for safe trajectory tracking. \eqref{cons:rate_cost} specifies the minimum effective control-link throughput required to achieve a prescribed LQG performance under the FBL regime, linking the reliability of control command delivery with the overall control performance. \eqref{cons:comm_rate} ensures that the communication link meets data transmission requirements in each time slot. Finally, \eqref{cons:resource_sum} and~\eqref{cons:alpha_range} jointly define the overall time-frequency resource budget and the admissible allocation set for sensing, communication, and control.

\subsection{Non-Convexity Analysis}

The joint resource allocation problem~\eqref{prob:resource_allocation} is inherently non-convex due to the strong and nested coupling among sensing, communication, and control in the ISCC closed-loop system.
First, the objective function~\eqref{prob:obj} depends on the steady-state tracking error covariance, which is the stabilizing solution to a modified algebraic Riccati equation whose coefficients jointly depend on the sensing accuracy and control-link reliability.
As a result, the objective function is only implicitly related to the resource allocation variables and is generally non-convex.
Second, the mean-square stability constraint~\eqref{cons:stability} involves a spectral radius condition on the state covariance evolution matrix.
Since the spectral radius is neither convex nor concave and depends on the control packet loss rate, which is itself a function of the control resource fraction, this constraint induces a non-convex feasible set.
Finally, the rate-cost constraint~\eqref{cons:rate_cost} couples the decoding success probability and the achievable transmission rate under the FBL regime, leading to a non-concave multiplicative structure.
These coupled properties render problem~\eqref{prob:resource_allocation} highly non-convex and preclude the direct application of standard convex optimization techniques.

\subsection{SCA-Based Resource Allocation Algorithm}\label{SCA}

Motivated by the non-convexity of problem~\eqref{prob:resource_allocation} and the lack of a closed-form solution, we develop a SCA-based algorithm to compute a stationary point.
By exploiting the threshold structure of the mean-square stability condition and the affine time-frequency resource budget, the original problem admits a low-dimensional reformulation with a simplified stability constraint. After this reformulation, the remaining non-convex components are smooth and can be efficiently handled via first-order convex approximation~\cite{scutari2016parallel}.

\subsubsection{Stability-Feasible Region Characterization}\label{Stability}

Before applying SCA, it is essential to explicitly characterize the feasibility region induced by the mean-square stability constraint~\eqref{cons:stability}, as the original spectral-radius condition $\rho(\mathbf{M})<1$ is non-convex and non-differentiable, making it difficult to incorporate directly into the optimization framework. To address this, we numerically evaluate the spectral radius as a function of the control packet loss rate and identify the maximum allowable packet loss that ensures stability.

\begin{figure*}[t]
    \centering
    \subfigure[]{
        \includegraphics[width=0.31\textwidth]{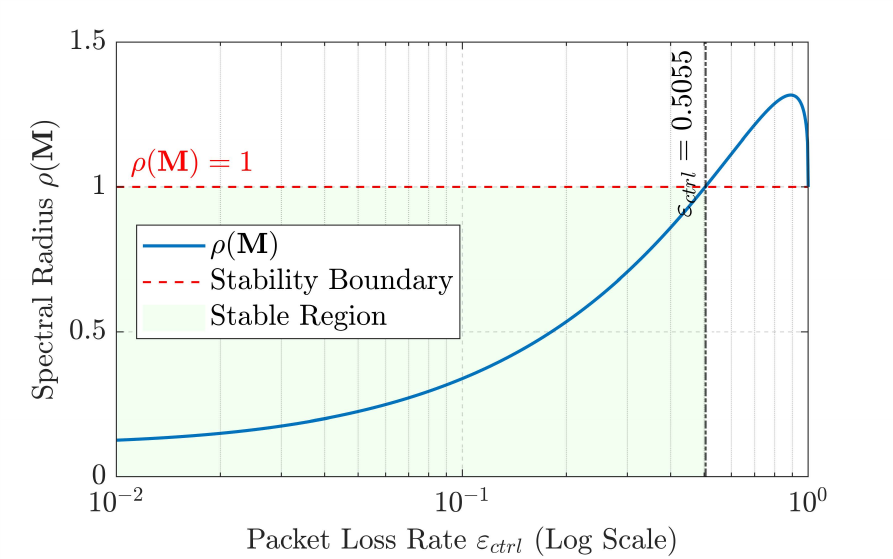}
        \label{fig:subfig1}
    }
    \subfigure[]{
        \includegraphics[width=0.31\textwidth]{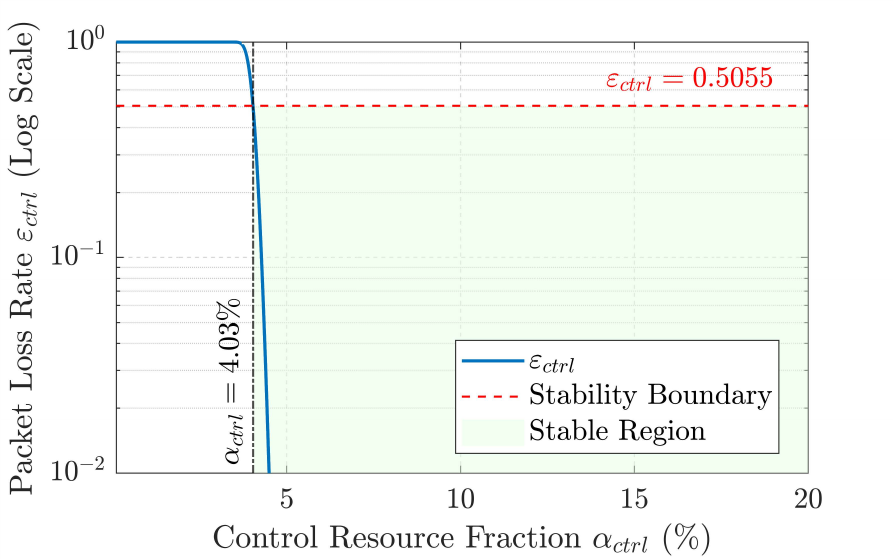}
        \label{fig:subfig2}
    }
    \subfigure[]{
        \includegraphics[width=0.31\textwidth]{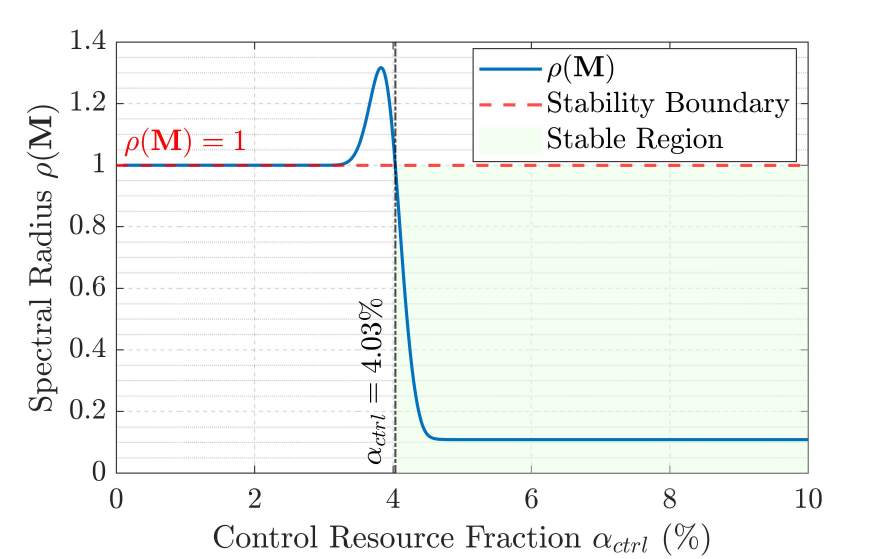}
        \label{fig:subfig3}
    }
    \caption{Stability and reliability characteristics of the control loop:
    (a) Spectral radius $\rho(\mathbf{M})$ of the closed-loop covariance evolution matrix versus the control packet loss rate $\varepsilon_{\rm ctrl}$;
    (b) Control packet loss rate $\varepsilon_{\rm ctrl}$ as a function of the control resource fraction $\alpha_{\rm ctrl}$;
    (c) Spectral radius $\rho(\mathbf{M})$ of the closed-loop covariance evolution matrix versus the control resource fraction $\alpha_{\mathrm{ctrl}}$.}
    \label{fig:combined}
\end{figure*}

Fig.~\ref{fig:subfig1} depicts the dependence of the spectral radius of the closed-loop covariance evolution matrix, $\rho(\mathbf{M})$, on the control packet loss rate, $\varepsilon_{\rm ctrl}$. Within the region $\rho(\mathbf{M}) \le 1$, the spectral radius increases monotonically with $\varepsilon_{\rm ctrl}$. When $\varepsilon_{\rm ctrl}$ reaches $\varepsilon_{\rm ctrl}^{\star} = 0.5055$, the system attains the mean-square stability boundary ($\rho(\mathbf{M}) = 1$), with $\varepsilon_{\rm ctrl}^{\star}$ defined as the maximum control packet loss rate allowed for stability. This critical value is determined by the drone dynamics, represented by $\mathbf{A}$ and $\mathbf{B}$, and the LQG feedback gain $\mathbf{K}$. While $\varepsilon_{\rm ctrl}^{\star}$ varies with system parameters, the presence of a stability threshold for control packet loss is a general feature of closed-loop systems. The relatively high value of $\varepsilon_{\rm ctrl}^{\star}$ indicates that the designed control link provides substantial stability margin.

Fig.~\ref{fig:subfig2} shows the variation of the control packet loss rate, $\varepsilon_{\rm ctrl}$, with the control resource fraction $\alpha_{\rm ctrl}$ computed based on the trajectory-averaged SNR. The maximum allowable control packet loss rate, $\varepsilon_{\rm ctrl}^{\star}$, corresponds to $\alpha_{\rm ctrl}^{\star} = 4.03\%$, where $\varepsilon_{\rm ctrl}$ exhibits a sharp transition. Once $\alpha_{\rm ctrl}$ exceeds this critical value, the packet loss rate rapidly drops below $\varepsilon_{\rm ctrl}^{\star}$, thereby ensuring mean-square stability. The threshold $\alpha_{\rm ctrl}^{\star}$, defined as the minimum control resource fraction required for stability, depends on the system configuration, including total transmit power, available bandwidth, and channel conditions.

Fig.~\ref{fig:subfig3} further shows the relationship between $\alpha_{\rm ctrl}$ and the closed-loop stability margin. When $\alpha_{\rm ctrl}=\alpha_{\rm ctrl}^{\star}$, $\rho(\mathbf{M})$ drops below one, marking the onset of mean-square stability. As $\alpha_{\rm ctrl}$ increases beyond approximately $5\%$, $\rho(\mathbf{M})$ saturates around $0.1$, indicating that additional control resource allocation provides negligible stability improvement.

Overall, Fig.~\ref{fig:combined} shows that the system is mean-square stable when the control resource fraction exceeds the critical threshold $\alpha_{\rm ctrl}^{\star}$, and unstable otherwise. Accordingly, the mean-square stability constraint $\rho(\mathbf{M})<1$ can be equivalently enforced by $\alpha_{\rm ctrl} > \alpha_{\rm ctrl}^{\star}$, avoiding the need to directly handle the spectral radius in the optimization process.

\subsubsection{Problem Reformulation and SCA-Based Solution}

With the stability-feasible region characterized, we reformulate problem~\eqref{prob:resource_allocation}. 
The affine resource constraint~\eqref{cons:resource_sum} allows elimination of the communication resource fraction as
$\alpha_{\mathrm{comm}} = 1 - \alpha_{\mathrm{sen}} - \alpha_{\mathrm{ctrl}}$,
reducing the optimization variables to
$\boldsymbol{\alpha} = [\alpha_{\mathrm{sen}}, \alpha_{\mathrm{ctrl}}]^{\mathsf T}$.
The communication rate constraint remains convex, while closed-loop stability is ensured by $\alpha_{\rm ctrl} > \alpha_{\rm ctrl}^{\star}$.

After this simplification, the sensing-dependent tracking error in the objective is convex, and the only non-convexity arises from the nonlinear control reliability term induced by FBL transmission. Both components are continuously differentiable, enabling SCA. Specifically, the objective $J(\boldsymbol{\alpha})$ is dominated by the sensing-induced PEB
$f_{\mathrm{sen}}(\alpha_{\mathrm{sen}}) \triangleq \mathrm{Tr}(\mathbf{CRLB}_\mathbf{p}(\alpha_{\mathrm{sen}}))$,
which is convex since CRLB scales inversely with resource fraction. Therefore, it can be minimized directly without linearization, preserving curvature and improving convergence. The control reliability constraint~\eqref{cons:rate_cost} involves
$g(\alpha_{\mathrm{ctrl}}) \triangleq (1-\varepsilon_{\mathrm{ctrl}}(\alpha_{\mathrm{ctrl}}))\mathcal{R}(\alpha_{\mathrm{ctrl}})$, for which a first-order affine lower bound at $\alpha_{\mathrm{ctrl}}^{(k)}$ is given by
\begin{equation}
\tilde{g}(\alpha_{\mathrm{ctrl}}; \alpha_{\mathrm{ctrl}}^{(k)})
=
g(\alpha_{\mathrm{ctrl}}^{(k)})
+
\nabla g(\alpha_{\mathrm{ctrl}}^{(k)})
\left(\alpha_{\mathrm{ctrl}} - \alpha_{\mathrm{ctrl}}^{(k)}\right).
\end{equation}

By substituting the original constraint with 
$\tilde{g}(\alpha_{\mathrm{ctrl}}; \alpha_{\mathrm{ctrl}}^{(k)}) \ge L(b)$ 
and adding a proximal term to the objective function for convergence, 
the convex sub-problem at iteration $k$ is formulated as
\begin{subequations}
\begin{align}
    \min_{\alpha_{\mathrm{sen}}, \alpha_{\mathrm{ctrl}}} \quad & f_{\mathrm{sen}}(\alpha_{\mathrm{sen}}) + \rho ||\boldsymbol{\alpha} - \boldsymbol{\alpha}^{(k)}||^2 \\
    \text{s.t.} \quad\,\,\,\, & \tilde{g}(\alpha_{\mathrm{ctrl}}; \alpha_{\mathrm{ctrl}}^{(k)}) \ge L(b), \\
    & \mathcal{R}_{\mathrm{comm}}(1 - \alpha_{\mathrm{sen}} - \alpha_{\mathrm{ctrl}}) \ge \mathcal{R}_{\mathrm{comm}}^{\min}, \\
    & \alpha_{\mathrm{sen}} + \alpha_{\mathrm{ctrl}} < 1, \\
    & 0 < \alpha_{i} < 1, \quad i \in \{\mathrm{sen}, \mathrm{ctrl}\}, \\
    & \alpha_{\mathrm{ctrl}} > \alpha_{\mathrm{ctrl}}^{\star}.
\end{align}
\end{subequations}
where $\rho>0$ is a penalty parameter that controls the update step, keeping the new solution within the trust region where the first-order approximation is accurate.

The above convex sub-problem is solved iteratively, with the linearization point updated at each iteration until convergence. The complete procedure is summarized in Algorithm~\ref{alg:sca_allocation}.

\begin{algorithm}[!t]
\caption{SCA-Based Resource Allocation for ISCC Systems}
\label{alg:sca_allocation}
\begin{algorithmic}[1]
\REQUIRE System parameters required to evaluate $f_{\mathrm{sen}}(\cdot)$ and $g(\cdot)$, and convergence tolerance $\delta$
\STATE Initialize $k=0$ and a feasible point $\boldsymbol{\alpha}^{(0)} = [\alpha_{\mathrm{sen}}^{(0)}, \alpha_{\mathrm{ctrl}}^{(0)}]^T$
\REPEAT
    \STATE Compute the value and gradient of $g(\cdot)$ at $\alpha_{\mathrm{ctrl}}^{(k)}$
    \STATE Construct first-order affine approximation of $g(\cdot)$ at $\alpha_{\mathrm{ctrl}}^{(k)}$
    \STATE Solve the resulting convex sub-problem to obtain $\boldsymbol{\alpha}^*$
    \STATE Update $\boldsymbol{\alpha}^{(k+1)} \leftarrow \boldsymbol{\alpha}^*$
    \STATE $k \leftarrow k+1$
\UNTIL $ \| \boldsymbol{\alpha}^{(k)} - \boldsymbol{\alpha}^{(k-1)} \|^2 \le \delta $
\ENSURE Stationary solution $\boldsymbol{\alpha}^*$
\end{algorithmic}
\end{algorithm}

Each SCA iteration solves a small convex program with only two variables ($n=2$) and simple linear constraints, resulting in negligible per-iteration cost. Hence, the overall complexity of Algorithm~\ref{alg:sca_allocation} scales linearly with the number of iterations $I_{\rm iter}$, i.e., $\mathcal{O}(I_{\rm iter})$, supporting quasi-real-time implementation.

\section{Performance Evaluation}\label{Performance Evaluation}

\subsection{Simulation Parameter Settings}

To evaluate the proposed ISCC-based drone trajectory tracking scheme, we consider a single-GBS scenario with a drone operating within the GBS coverage. System parameters are summarized in Table~\ref{simulation}. In the simulation, the GBS is located at the origin, and its coverage is modeled as an upper hemisphere with a radius of $1000\,\mathrm{m}$. The drone is assumed to remain within this coverage throughout the flight. The reference flight trajectory is generated by randomly placing key waypoints according to a 3D homogeneous Poisson point process, subject to altitude and coverage constraints, to ensure spatial randomness and representativeness. A continuous and smooth trajectory is then constructed via cubic spline interpolation through these waypoints. All numerical results are averaged over multiple realizations of the randomly generated trajectories to avoid dependence on any specific flight path.

\begin{table}[!t]
\caption{The Main Simulation Parameters}\label{simulation}
\centering
\begin{tabular}{|l|l|}
  \hline
  \textbf{Parameters} & \textbf{Value}\tabularnewline
  \hline
  \hline
  Carrier frequency ${f_c}$ & 2.4\,GHz~\cite{3gpp.38.331} \\
  \hline
  System bandwidth $B$ & 10\,MHz~\cite{3gpp.38.331} \\
  \hline
  Transmission power of GBS & 42\,dBm \\
  \hline
  GBS antenna height & 25~m \\
  \hline
  BS antenna configuration ${N_x} \times {N_y}$ & 2 $\times$ 4 \\
  \hline
  BS coverage radius  & 1\,km \\
  \hline
  Drone altitude range &  (100, 300]\,m~\cite{3gpp.38.331} \\
  \hline
  Drone antenna configuration & Omni-directional \\
  \hline
  Shadow fading $\sigma _{{\rm{SF}}}$ & $4.64{e^{ - 0.0066{h}}}$\,dB~\cite{3gpp.38.331} \\
  \hline
  RCS $\sigma _{\rm RCS}$ & 0.01\,m$^2$~\cite{zhang2025channel} \\
  \hline
  Noise PSD $N_0$ & -150\,dBm/Hz \\
  \hline
  Number of OFDM symbols  \({\cal M}\) & 64~\cite{gaudio2019performance} \\
  \hline
  Number of subcarriers  \({\cal K}\) & 100 \\
  \hline
  Subcarrier spacing $\Delta f$ & 100\,kHz \\
  \hline
  Symbol duration $T_{\text{sym}}$ & 12.5\,$\mu s$ \\
  \hline
  Flight duration $T_{\mathrm{m}}$ & $500\,\mathrm{s}$ \\
  \hline
  Time solt $\Delta t$ & $0.1\,\mathrm{s}$ \\
  \hline
  LQG state weight for position error $\mathbf{Q}_\mathbf{p}$ & $10^{4}$ \\
  \hline
  LQG state weight for velocity error $\mathbf{Q}_\mathbf{v}$ & $10^{2}$ \\
  \hline
  LQG control weight matrix $\mathbf{R}$ & $0.01\,\mathbf{I}_3$ \\
  \hline
  Control command size $S$ & 256\,bits \\
  \hline
  Control blocklength $L_{\mathrm{ctrl}}$ & 50~symbols \\
  \hline
  Data blocklength $L_{\mathrm{comm}}$ & 8000~symbols \\
  \hline
  BLER for control signaling $\varepsilon_{\mathrm{ctrl}}$ & $10^{-5}$~\cite{sutton2019enabling} \\
  \hline
  BLER for data communication $\varepsilon_{\mathrm{comm}}$ & $10^{-3}$~\cite{sutton2019enabling} \\
  \hline
  Position process noise variance & 0.2~m$^2$ \\
  \hline
   Velocity process noise variance & 0.002~(m/s)$^2$ \\
    \hline

\end{tabular}
\end{table}

\subsection{Simulation Analysis}

\subsubsection{Impact of Resource Allocation on the LQG Cost}

\begin{figure*}[t]
    \centering
    \subfigure[]{
        \includegraphics[width=0.31 \textwidth]{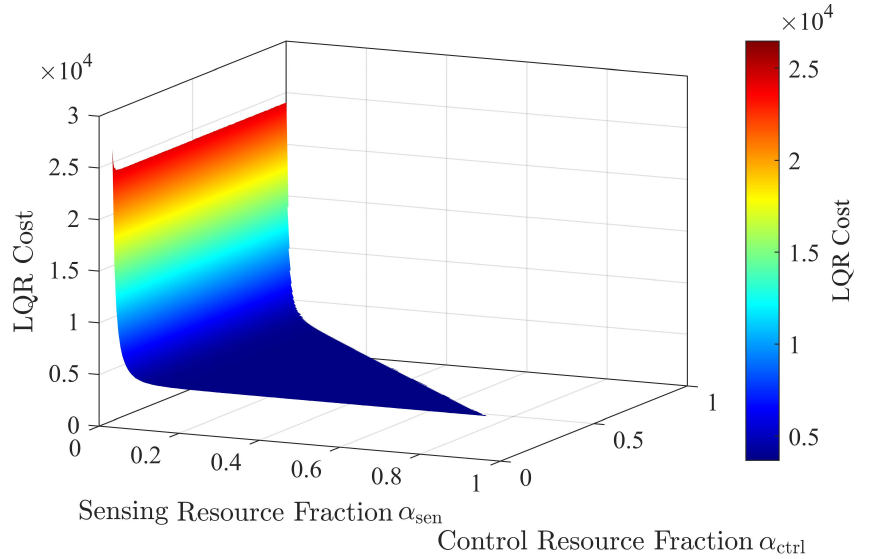}
        \label{fig:LQG0}
    }
    \subfigure[]{
        \includegraphics[width=0.31 \textwidth]{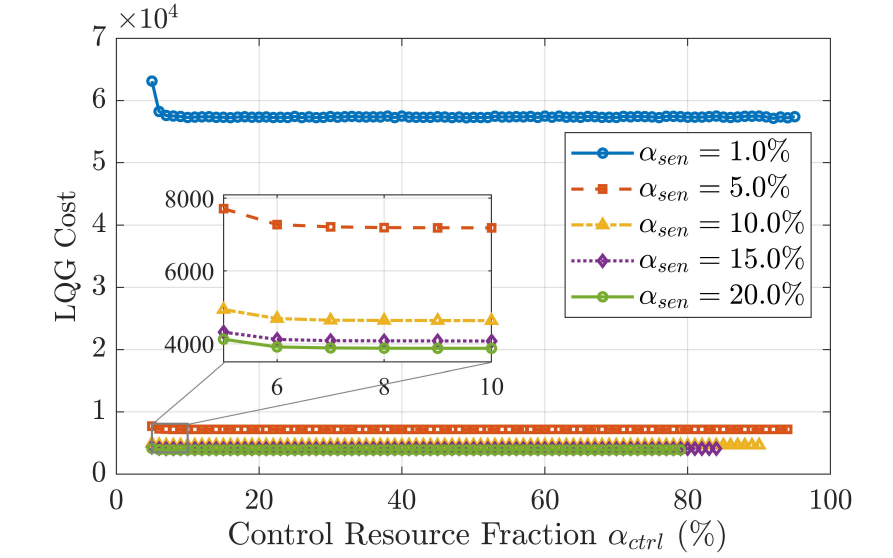}
        \label{fig:LQG1}
    }
    \subfigure[]{
        \includegraphics[width=0.31 \textwidth]{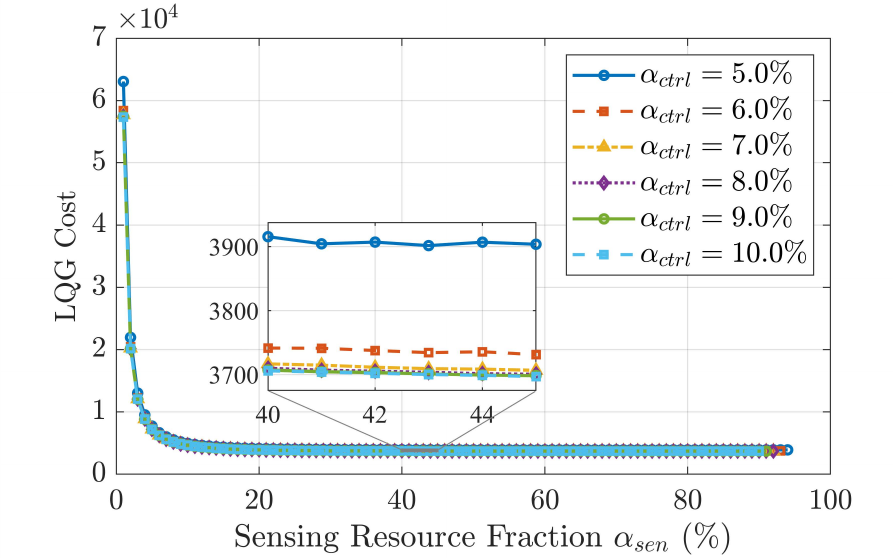}
        \label{fig:LQG2}
    }
    \caption{Impact of sensing and control resource allocation on the long-term average LQG cost in the ISCC closed-loop system. 
(a) LQG cost versus control resource fraction $\alpha_{\mathrm{ctrl}}$ and sensing resource fraction $\alpha_{\mathrm{sen}}$. 
(b) LQG cost versus control resource fraction $\alpha_{\mathrm{ctrl}}$ for different sensing resource fractions $\alpha_{\mathrm{sen}}$. 
(c) LQG cost versus sensing resource fraction $\alpha_{\mathrm{sen}}$ for different control resource fractions $\alpha_{\mathrm{ctrl}}$.}
    \label{fig:LQG_combined}
\end{figure*}

Fig.~\ref{fig:LQG_combined} shows the long-term average LQG cost of the system under different sensing and control time-frequency resource fractions, \( \alpha_{\mathrm{sen}} \) and \( \alpha_{\mathrm{ctrl}} \). Only results with the control resource fraction above the mean-square stability threshold are shown to avoid masking performance trends caused by divergence in unstable regions. 
As shown in Fig.~\ref{fig:LQG0}, the LQG cost decreases monotonically as the sensing resource fraction \( \alpha_{\mathrm{sen}} \) increases. In contrast, it is much less sensitive to variations in the control resource fraction \( \alpha_{\mathrm{ctrl}} \), remaining largely flat along this dimension. To further characterize the impact of the two types of resources, Fig.~\ref{fig:LQG1} and  Fig.~\ref{fig:LQG2} present the corresponding cross-sectional results with fixed \( \alpha_{\mathrm{sen}} \) and fixed \( \alpha_{\mathrm{ctrl}} \), respectively. It can be observed that moderately increasing the control resource fraction \( \alpha_{\mathrm{ctrl}} \) (approximately \(5\%\)-\(7\%\)) helps reduce the LQG cost, with a more noticeable effect when the sensing resource is low (e.g., \( \alpha_{\mathrm{sen}} = 1\% \)). However, once \( \alpha_{\mathrm{ctrl}} \) exceeds approximately \(7\%\), further increasing \( \alpha_{\mathrm{ctrl}} \) yields no significant improvement in the LQG cost. By comparison, \( \alpha_{\mathrm{sen}} \) has a more pronounced impact on the LQG cost. As \( \alpha_{\mathrm{sen}} \) increases, the LQG cost continuously decreases and gradually flattens when \( \alpha_{\mathrm{sen}} \) exceeds approximately \(15\%\), eventually stabilizing at around \( 3.7 \times 10^{3} \). 
Overall, the control resource must exceed the critical threshold $\alpha_{\rm ctrl}^{\star}$ in Section~\ref{Stability} to ensure system stability. Beyond this threshold, its impact is limited, and the closed-loop tracking performance is primarily governed by the sensing resource.

\subsubsection{Impact of Resource Allocation on Trajectory Tracking Error}

\begin{figure*}[t]
    \centering
    \subfigure[]{
        \includegraphics[width=0.31 \textwidth]{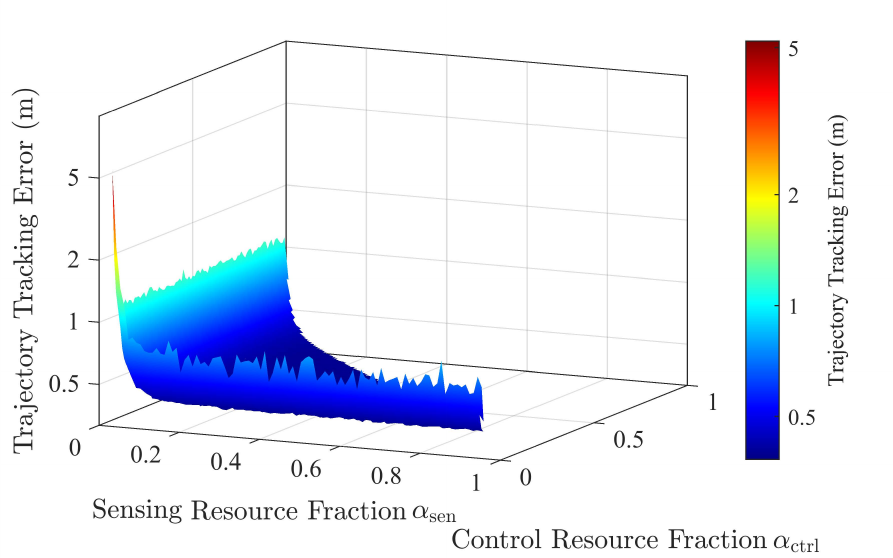}
        \label{fig:error0}
    }
    \subfigure[]{
        \includegraphics[width=0.31 \textwidth]{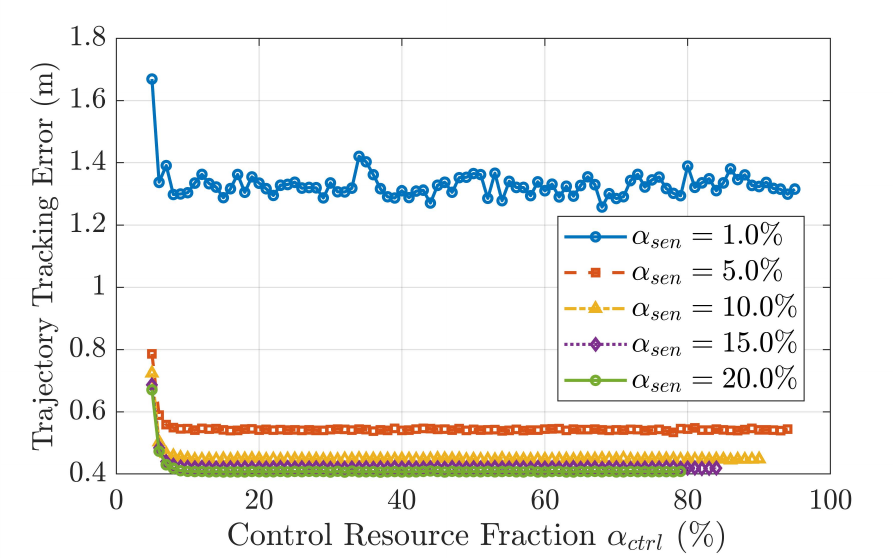}
        \label{fig:error1}
    }
    \subfigure[]{
        \includegraphics[width=0.31 \textwidth]{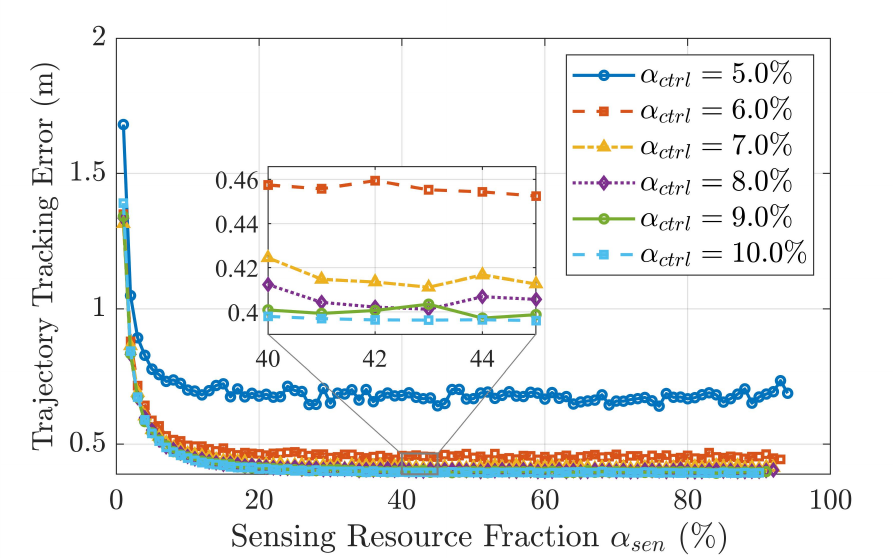}
        \label{fig:error2}
    }
\caption{Impact of sensing and control resource allocation on trajectory tracking error in the ISCC closed-loop system. 
(a) Tracking error versus control resource fraction $\alpha_{\mathrm{ctrl}}$ and sensing resource fraction $\alpha_{\mathrm{sen}}$. 
(b) Tracking error versus control resource fraction $\alpha_{\mathrm{ctrl}}$ for different sensing resource fractions $\alpha_{\mathrm{sen}}$. 
(c) Tracking error versus sensing resource fraction $\alpha_{\mathrm{sen}}$ for different control resource fractions $\alpha_{\mathrm{ctrl}}$.}

    \label{fig:error_combined}
\end{figure*}

Fig.~\ref{fig:error_combined} shows the impact of the sensing and the control time-frequency resource fractions, \( \alpha_{\mathrm{sen}} \) and \( \alpha_{\mathrm{ctrl}} \), on the trajectory tracking error.
As shown in Fig.~\ref{fig:error0}, the trajectory tracking error becomes significantly large when either the sensing or control resource fraction is insufficient, and reaches pronounced peaks when both are low. As \( \alpha_{\mathrm{sen}} \) and \( \alpha_{\mathrm{ctrl}} \) increase, the trajectory tracking error gradually decreases and eventually stabilizes, forming a relatively flat stable operating region.
Fig.~\ref{fig:error1} and Fig.~\ref{fig:error2} further present the cross-sectional results obtained under fixed sensing resource fraction \( \alpha_{\mathrm{sen}} \) and fixed control resource fraction \( \alpha_{\mathrm{ctrl}} \), respectively. As \( \alpha_{\mathrm{ctrl}} \) increases from \(5\%\) to \(6\%\), the trajectory tracking error decreases rapidly; however, when \( \alpha_{\mathrm{ctrl}} \) exceeds approximately \(7\%\), the performance improvement tends to saturate. In contrast, under sufficiently large \( \alpha_{\mathrm{ctrl}} \), increasing \( \alpha_{\mathrm{sen}} \) can further reduce the trajectory tracking error, but the marginal gain becomes less pronounced when \( \alpha_{\mathrm{sen}} \) exceeds approximately \(10\%\), with the tracking error eventually stabilizing at around \(0.4\)~m.
Overall, Fig.~\ref{fig:error_combined} indicates that, once closed-loop stability is guaranteed, further increasing the control resource yields marginal performance improvement, whereas the sensing resource predominantly determines the achievable tracking accuracy within the stable region.

\subsubsection{Relationship Between Trajectory Tracking Error and Sensing Error}

\begin{figure}[t]
    \centering
    \includegraphics[width=3.0in]{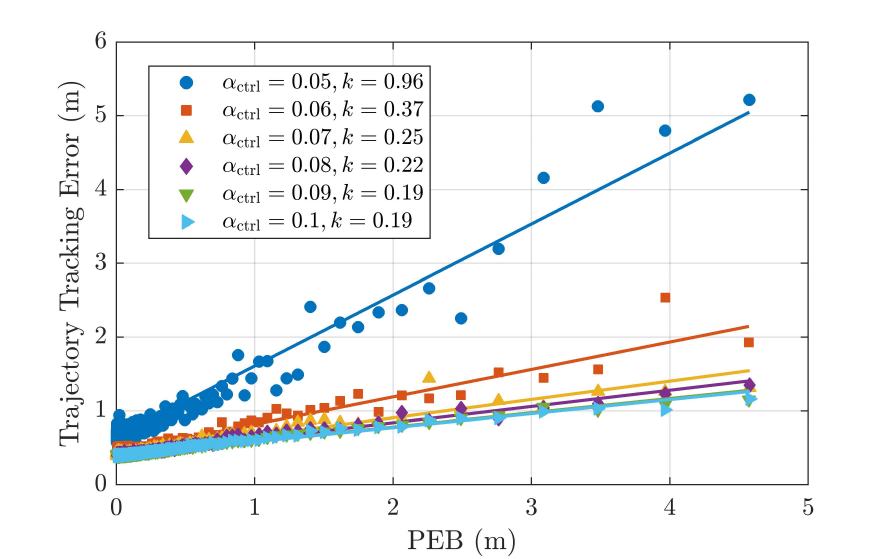}
    \caption{Average trajectory tracking error as a function of the PEB for different control resource fractions $\alpha_{\mathrm{ctrl}}$.}
    \label{tu14}
\end{figure}

Fig.~\ref{tu14} shows the average trajectory tracking error as a function of the PEB under different control resource fractions \( \alpha_{\mathrm{ctrl}} \), where the solid lines represent linear fits and the corresponding slopes are denoted by \( k \). It can be observed that, for all considered values of \( \alpha_{\mathrm{ctrl}} \), the trajectory tracking error exhibits an approximately linear dependence on the PEB, indicating that sensing errors are propagated to the tracking performance through the closed-loop feedback.
As \( \alpha_{\mathrm{ctrl}} \) increases, the fitted slope \( k \) gradually decreases, indicating that the closed-loop system becomes less sensitive to sensing inaccuracy as the reliability of control command transmission improves. When \( \alpha_{\mathrm{ctrl}} \) reaches relatively high levels (e.g., \(9\%\) and \(10\%\)), the values of \( k \) become nearly identical, suggesting that further increasing \( \alpha_{\mathrm{ctrl}} \) provides limited additional reduction in sensing-error-induced performance degradation.
We observe that the fitted slopes satisfy \( k < 1 \) for all considered values of \( \alpha_{\mathrm{ctrl}} \), indicating that sensing errors are attenuated rather than amplified through the closed-loop dynamics under the proposed ISCC framework.

\subsubsection{Coupling Among Sensing, Communication, and Control Performance and Its Impact on Trajectory Tracking Error}

\begin{figure}[t]
    \centering
    \includegraphics[width=3.0in]{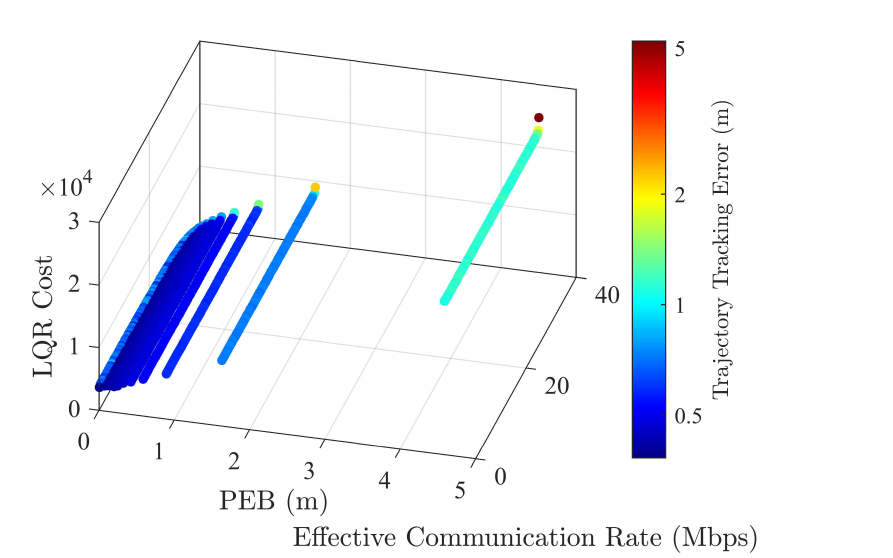}
    \caption{Joint relationship among sensing error, effective communication rate, and LQG cost, and their impact on the average trajectory tracking error.}
    \label{tu15}
\end{figure}

Fig.~\ref{tu15} shows the coupling among communication, sensing, and control performances and its impact on the trajectory tracking error. 
First, from the perspective of the coupling between sensing and control, the LQG cost increases linearly with the PEB, leading to higher trajectory errors.
Second, the communication dimension reveals the resource competition effect. Within a moderate communication rate range, the data points concentrate in regions of low PEB and low LQG cost, indicating that trajectory tracking errors remain small. However, when the communication rate exceeds approximately 35~Mbps, the system is forced to allocate excessive resources to the data link, which encroaches upon the sensing and control resources ($\alpha_{\rm sen}$ and $\alpha_{\rm ctrl}$), causing PEB, LQG cost, and trajectory tracking errors to rise sharply.
In summary, under resource-limited conditions, a resource-driven ISCC closed-loop error propagation chain emerges: competition for communication resources reduces sensing accuracy (high PEB), which in turn increases the control cost (high LQG cost) and ultimately amplifies trajectory tracking errors.

\subsubsection{Performance Comparison of Different Resource Allocation Strategies}\label{Resource Allocation}

Table~\ref{tab:comparison} compares the proposed SCA-based Resource allocation strategy with the benchmark grid search (Grid)~\cite{shams2024water}, genetic algorithm (GA)~\cite{alhijawi2024genetic}. 
The joint optimization problem in~\eqref{cons:alpha_range} involves highly nonlinear couplings among sensing, communication, and control metrics, making it strongly non-convex and difficult to solve globally. To obtain a performance upper bound, the Grid method is used to approximate the global optimum. First, SCA achieves performance close to Grid in terms of tracking error, PEB, and LQG cost, indicating that its first-order Taylor approximation accurately captures the local geometry of the non-convex constraints and converges to a stationary point near the global optimum. Although GA achieves slightly lower tracking error, the difference is only at the millimeter level.
Second, in terms of transmission rate, SCA attains an average rate of 28.15~Mbps, which is higher than both Grid and GA. Under the current setup, the theoretical maximum transmission rate of the system's communication link is 35~Mbps. Finally, regarding computational efficiency, SCA requires only 68.6~s of runtime, significantly lower than the other methods. Overall, the proposed SCA method achieves a trajectory tracking error of 0.4122~m, with only a millimeter-level difference compared to the Grid benchmark, while requiring approximately $0.68\%$ of its computation time.

\begin{table}[htbp]
\centering
\caption{Comparison of Different Resource Allocation Strategies}
\label{tab:comparison}
\renewcommand{\arraystretch}{1.1}
\setlength{\tabcolsep}{4pt}
\begin{tabular}{llll}
\toprule
\textbf{Strategy} 
& \textbf{SCA} & \textbf{Grid} & \textbf{GA} \\
\midrule
\multicolumn{4}{l}{\textit{\textbf{Performance Metrics}}} \\
\midrule
Tracking error (m) 
& 0.4122 & 0.4077 & 0.4109  \\
Transmission rate (Mbps) 
& 28.15 & 25.33 & 25.40  \\
PEB (m) 
& 0.0635 & 0.0314 & 0.0386 \\
LQG cost 
& 3729.46 & 3690.14 & 3696.54 \\
\midrule
\multicolumn{4}{l}{\textit{\textbf{Computational Efficiency}}} \\
\midrule
Time (s) 
& 68.60 & 10154.78 & 694.84 \\
\midrule
\multicolumn{4}{l}{\textit{\textbf{Resource Allocation}}} \\
\midrule
Sensing resource fraction (\%) 
& 20 & 24 & 21 \\
Control resource fraction (\%) 
& 20 & 13 & 16 \\
Communication resource fraction (\%)  
& 60 & 63 & 63 \\
\bottomrule
\end{tabular}
\end{table}

\subsubsection{Trajectory Tracking Performance Comparison of Different Schemes}

Based on the optimal resource allocation obtained in Section~\ref{Resource Allocation}, the trajectory tracking performance of the proposed ISCC closed-loop scheme is evaluated and compared with the following benchmark schemes:
i) Open-loop scheme: a purely feedforward control strategy that executes predefined acceleration commands according to the reference trajectory without incorporating any state feedback.
ii) GNSS-based closed-loop scheme: a feedback control scheme relying on GNSS observations corrupted by zero-mean Gaussian noise, where the standard deviations of the position and velocity measurements are set to $3$~m and $0.2$~m/s, respectively~\cite{beard2012small}.
iii) ISCC ignoring control packet loss: an ISCC scheme designed under the assumption of error-free control command delivery, whose resulting resource allocation is evaluated in a practical system subject to FBL-induced packet losses~\cite{lei2023control, lei2024edge}.

\begin{figure}[t]
    \centering
    \subfigure[]{
        \includegraphics[width=0.4\textwidth]{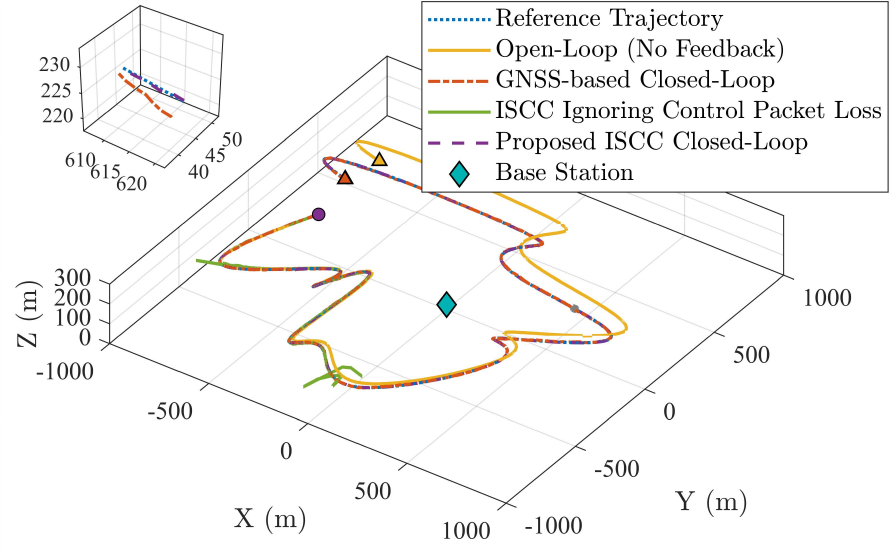}
        \label{fig:Error1}
    }
    \hspace{0.5cm}
    \subfigure[]{
        \includegraphics[width=0.4\textwidth]{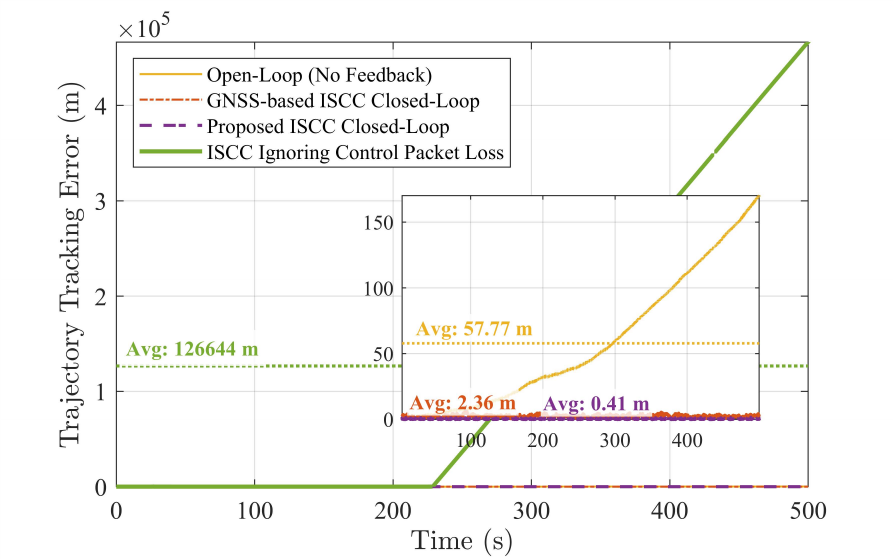}
        \label{fig:Error2}
    }
    \caption{Trajectory tracking performance comparison of a drone under different control schemes:
    (a) 3D flight trajectories;
    (b) trajectory tracking error versus time.}
    \label{fig:Error}
\end{figure}

Fig.~\ref{fig:Error1} shows the 3D flight trajectories under different control schemes.
The open-loop scheme exhibits progressive deviation from the reference trajectory due to the absence of feedback.
The GNSS-based closed-loop scheme partially mitigates this deviation, yet noticeable tracking errors remain along the flight path.
The ISCC ignoring control packet loss scheme shows severe trajectory divergence after a certain time.
In contrast, the proposed ISCC closed-loop scheme closely follows the reference trajectory throughout the mission.
Fig.~\ref{fig:Error2} presents the time evolution of the trajectory tracking error.
The open-loop scheme experiences rapid error growth, resulting in an average tracking error of approximately $57.8$~m.
With feedback enabled, the GNSS-based closed-loop scheme stabilizes the tracking error at a lower level, achieving an average error of about $2.36$~m.
The ISCC ignoring control packet loss scheme fails to maintain closed-loop stability, causing the trajectory tracking error to diverge numerically to an average of $1.2 \times 10^5$ m over the mission horizon.
By comparison, the proposed ISCC closed-loop scheme consistently maintains the smallest tracking error, with an average value of $0.41$~m over the entire horizon.

From Fig.~\ref{fig:Error1} and Fig.~\ref{fig:Error2}, we observe that ISCC resource allocation designs that ignore control packet loss may lead to closed-loop instability in the presence of FBL-induced transmission impairments. By explicitly accounting for control command unreliability, the proposed SC$^2$ closed-loop scheme achieves robust trajectory tracking, with an average error reduced to $17.37\%$ of that of the GNSS scheme.

\section{Conclusion}\label{Conclusion}
To address the lack of accurate closed-loop modeling of sensing errors and wireless control-link unreliability in LAWN scenarios, this paper developed an ISCC closed-loop framework for drone trajectory tracking in LAWN scenarios, explicitly incorporating ISAC sensing errors and stochastic control packet losses into the system dynamics. The framework unifies the coupling among communication rate, sensing accuracy, and control reliability, and reveals the distinct roles of time-frequency and power resources: sensing resources suppress observation noise and determine state estimation accuracy, while control resources ensure reliable command delivery and govern mean-square stability. A lower bound on control resource allocation to guarantee stability was derived, and a time-frequency resource allocation strategy was formulated to minimize trajectory tracking error under communication and stability constraints, solved via SCA. Simulations show that, once stability is ensured, system performance is dominated by sensing accuracy, with tracking error approximately linearly dependent on sensing error. The proposed ISCC scheme achieved an average tracking error of $0.41$~m, only $17.37\%$ of that of a GNSS-based baseline, while avoiding divergence under FBL transmission, demonstrating significant performance gains in LAWN scenarios.

\bibliographystyle{ieeetr}
\bibliography{ref}

\end{document}